\documentclass[11pt,a4paper]{article}
\pdftrailerid{}
\usepackage[T1]{fontenc}
\usepackage[utf8]{inputenc}
\usepackage{lmodern}
\usepackage{mathtools,amssymb,mathrsfs}
\usepackage{setspace}
\usepackage[margin=26mm]{geometry}
\usepackage{microtype}
\usepackage{ragged2e}
\usepackage{parskip}
\usepackage{graphicx,subcaption,float}
\usepackage{needspace}
\usepackage{placeins}
\usepackage{etoolbox}
\usepackage{booktabs}
\usepackage{array}
\usepackage{xcolor}
\usepackage{xurl}
\usepackage[unicode,hidelinks,hypertexnames=false]{hyperref}
\usepackage{bookmark}
\makeatletter
\newcommand{\labelunnumbered}[1]{\begingroup\def\@currentlabel{}\label{#1}\endgroup}
\makeatother
\usepackage{cite} 
\AtBeginEnvironment{equation}{\Needspace{4\baselineskip}}
\hypersetup{
  pdftitle={Pfaffian sensitivity to temporal deformations of an N=5 background in the Lorentzian IIB matrix model},
  pdfauthor={Tetsuyuki Muramatsu},
  pdfcreator={LaTeX}}
\begin{document}
\thispagestyle{plain}
\vspace*{4mm}
\begin{center}
{\LARGE\bfseries Pfaffian sensitivity to temporal deformations\\[0.15em]
of an \(\boldsymbol{N=5}\) background in the\\[0.15em]
Lorentzian IIB matrix model\par}
\vspace{23mm}
{\normalsize Tetsuyuki Muramatsu\footnotemark\par}
\vspace{4mm}
{\normalsize Tokai National Higher Education and Research System\par}
\vspace{1mm}
{\normalsize Furo-cho, Chikusa-ku, Nagoya, 464-8601, Japan\par}
\end{center}
\vspace{17mm}
\footnotetext{The views expressed in this paper are solely those of the author and do not necessarily reflect those of the affiliated organization.}
\begin{abstract}\normalsize\setstretch{1.12}
How spacetime dimensionality and anisotropy emerge from matrix degrees of freedom is a central question in the IIB matrix model. Our previous work constructed a family of constrained classical solutions at \(N=5\), including a reference configuration with vanishing temporal extent and a pre-existing \(3+6\) spatial hierarchy. Here we examine how the fermionic Pfaffian responds to temporal-matrix deformations of this configuration.

The background geometry determines a distinguished real two-plane of temporal deformations. Numerical analysis of the full fermionic operator shows that the sensitivity of its lowest positive-eigenvalue doublet is strongly concentrated in this plane and dominates the first-order Pfaffian response. Under stated reflection-covariance and nonsingularity conditions, the other local one-loop contributions cancel at first order when combined, so the gradients of the logarithmic magnitudes of the total local one-loop weight and the Pfaffian agree.

We prescribe a temporal deformation along the projection of this gradient onto the two-plane and determine the remaining matrix components by conditional classical stationarity. The existing spatial hierarchy is modestly strengthened at quadratic order, while the spectrum changes little at the small amplitude considered. This quadratic response is independent of orientation within the plane, although the first-order Pfaffian response distinguishes such orientations. These results connect background geometry, fermionic spectral sensitivity, and conditional classical spatial anisotropy.
\end{abstract}

\clearpage
\setstretch{1.12}
\setlength{\parskip}{0pt}
\setlength{\parindent}{1.5em}

\section{Introduction}\label{sec:intro}

The emergence of spacetime dimensionality and anisotropy from matrix degrees of freedom is a central question in the IIB matrix model, proposed as a nonperturbative formulation of type IIB superstring theory \cite{ishibashi1997,aoki1998}. Integrating out its Majorana--Weyl fermions yields a Pfaffian that contributes to the quantum weight. We investigate how this contribution responds to temporal-matrix deformations around an explicit finite-\(N\) Lorentzian background.

In the Euclidean model, the role of the complex Pfaffian phase in rotational-symmetry breaking has been studied \cite{nishimuraVernizzi2000,anagnostopoulos2020}. Lorentzian investigations have addressed expanding spacetimes, anisotropy, classical backgrounds, and one-loop effective actions \cite{kim2012,nishimuraTsuchiya2019,anagnostopoulos2023,anagnostopoulos2026,hatakeyama2020,liao2026,steinacker2023}, while related one-loop analyses have considered noncommutative-brane and cosmological backgrounds \cite{blaschkeSteinacker2011,battistaSteinacker2024}. Our focus is the local directional response of the full fermionic operator around a constrained finite-\(N\) background.

Our previous paper \cite{muramatsu2026} constructed a constrained stationary solution family at \(N=5\), its physical modulus, and the associated local one-loop structure. A non-exhaustive search identified at least 23 stationary-solution classes, four of which admitted numerically continued one-parameter local families. We choose a previously analyzed branch with vanishing temporal extent in an appropriate Lorentz rest frame and a pre-existing \(3+6\) spatial hierarchy. Its regular local continuation and previously characterized fluctuation properties make it a suitable reference for examining how the Pfaffian responds to temporal deformations and how the existing spatial anisotropy changes.

An exact nearby constrained stationary solution admits an analytic local modulus, represented by a phase-parameterized family that preserves the quadratic background geometry. Rigorous pointwise bounds relate 23 numerical configurations sampled along the selected branch to this analytic family, after color-gauge and connected Lorentz transformations determined individually for each configuration. These classical results support the background construction; the fermionic and one-loop quantities below are evaluated at the numerical reference background.

After removing Lorentz-frame directions, we construct a distinguished real two-plane of temporal deformations from the internal geometry of the background. Numerical analysis of the full fermionic operator shows that the first-order sensitivity of its lowest positive-eigenvalue doublet is strongly concentrated in this plane and that the doublet dominates the Pfaffian gradient. A color--spin coupling vector relates this sensitivity to the temporal Hessian.

The local one-loop density is defined with the temporal-deformation coefficients and the classical modulus retained. In addition to the fermionic Pfaffian, it contains residual bosonic, gauge-fixing, and measure factors. Under the stated reflection-covariance and nonsingularity conditions, the combined non-Pfaffian contribution to the first derivative of the logarithmic magnitude vanishes at the reference point. The gradients of the logarithmic magnitudes of the total local one-loop weight and the Pfaffian therefore agree in the temporal directions. We prescribe a deformation along the projection of this gradient onto the two-plane and determine the remaining matrix components by classical stationarity. The resulting conditional quadratic response modestly strengthens the pre-existing \(3+6\) spatial hierarchy.

Section~\ref{sec:model} specifies the model and reference background, and Section~\ref{sec:temporal-extent} examines the classical temporal extent under deformations of \(A_0\). Sections~\ref{sec:geometry} and~\ref{sec:mechanism} develop the geometric two-plane and the fermionic response, while Section~\ref{sec:spatial} analyzes the conditional spatial response. Section~\ref{sec:discussion} discusses the conclusions. The appendices supply the fluctuation and fermionic derivations and the exact classical construction.

\section{Model, reference background, and temporal matrix deformations}\label{sec:model}

This section specifies the constrained finite-\(N=5\) reference background constructed in our previous paper \cite{muramatsu2026}, its temporal and spatial extents, and the independent directions used to deform \(A_0\).

\subsection{Lorentzian IIB matrix model and fixed-scale classical solutions}\label{sec:model-scale}

The fundamental variables of the IIB matrix model are ten Hermitian \(N\times N\) matrices \(A_\mu\) (\(\mu=0,\ldots,9\)) and a 16-component Majorana--Weyl fermionic matrix \(\psi^\alpha\). We remove the trace \(U(1)\) components, which decouple from the interactions, and work with traceless matrices. Here ``color'' refers to the \(N\)-dimensional matrix-index space, not spacetime. For \(N=5\), each bosonic matrix has 24 real degrees of freedom, and the fermions have 384 Grassmann components in total.

We use the mostly-plus spacetime metric \(\eta_{\mu\nu}=\operatorname{diag}(-1,+1,\ldots,+1)\). Lorentz indices are raised and lowered with this metric, and repeated Lorentz indices are summed according to the Einstein convention.

The action of the model is

\begin{equation}
S_{\rm IIB}[A,\psi]
=\frac{1}{g^2}\operatorname{Tr}\left[
\frac14F_{\mu\nu}F^{\mu\nu}
+\frac{i}{2}\psi^\alpha\gamma^\mu_{\alpha\beta}D_\mu^A\psi^\beta\right].
\label{eq:action}
\end{equation} Here \(g\) is the coupling constant, and \(F_{\mu\nu}\) is defined by the commutator of the bosonic matrices,

\begin{equation}
F_{\mu\nu}:=i[A_\mu,A_\nu].
\label{eq:field-strength}
\end{equation} The operator \(D_\mu^A\) acts in the adjoint representation and is defined by

\begin{equation}
D_\mu^AX:=i[A_\mu,X].
\label{eq:adjoint-derivative}
\end{equation}

The fermion \(\psi^\alpha\) is a ten-dimensional Majorana--Weyl spinor of positive chirality, with components labeled by \(\alpha=1,\ldots,16\). The matrices \(\gamma^\mu_{\alpha\beta}\) in the action define the chiral-spinor bilinear with charge conjugation incorporated. We use the same gamma-matrix and spinor-basis conventions as in our previous paper \cite{muramatsu2026}. The fermionic operator used below is defined in Section~\ref{sec:oneloop}.

We retain the Lorentzian oscillatory weight \(\exp(iS_{\rm IIB})\). The local bosonic Gaussian is defined by an infinitesimal positive-definite damping limit. The global integration-cycle question is separate; Euclidean--Lorentzian inequivalence has been reported in a fermion-omitted \(N=2\) model \cite{chou2025}.

We study classical configurations with vanishing background fermions. We write the quartic quantity obtained by removing the coupling factor \(g^{-2}\) from the bosonic action as

\begin{equation}
S_{B,0}[A]:=\frac14\operatorname{Tr}(F_{\mu\nu}F^{\mu\nu}).
\label{eq:boson-action}
\end{equation} The bosonic action is \(S_B[A]=S_{B,0}[A]/g^2\).

To compare the shapes and deformations of classical solutions at a common scale, we use the Lorentz-invariant quadratic quantity

\begin{equation}
P[A]:=\operatorname{Tr}(A_\mu A^\mu).
\label{eq:scale-P}
\end{equation} This contraction uses the Lorentzian metric and can take either sign for a general matrix configuration.

Following the conventions of our previous paper \cite{muramatsu2026}, we define

\begin{equation}
Q_\rho[A]:=P[A]-(6+2\sqrt3)
\label{eq:scale-Q}
\end{equation} and fix

\begin{equation}
Q_\rho[A]=9.
\label{eq:fixed-Q}
\end{equation} The classical backgrounds considered here therefore satisfy

\begin{equation}
P[A]=q_0,\qquad q_0:=15+2\sqrt3.
\label{eq:fixed-P}
\end{equation} The choice \(Q_\rho=9\) fixes a common normalization; it is not a dynamical scale selected by the classical equations.

Classical stationary configurations at fixed scale are obtained as stationary points of the function

\begin{equation}
\mathcal L_Q[A,\lambda_Q]
:=S_{B,0}[A]-\lambda_Q\bigl(Q_\rho[A]-9\bigr).
\label{eq:lagrangian-Q}
\end{equation} Here \(\lambda_Q\) is the Lagrange multiplier imposing the fixed-scale condition and is determined together with the stationary configuration. Since we have removed the coupling constant from the action, the corresponding multiplier for the original bosonic action \(S_B\) is \(\lambda_Q/g^2\).

Varying Eq.~\eqref{eq:lagrangian-Q} with respect to \(A_\mu\) gives

\begin{equation}
[A^\nu,[A_\nu,A_\mu]]=2\lambda_QA_\mu.
\label{eq:constrained-EOM}
\end{equation} The constrained classical solutions considered here satisfy both this equation and the fixed condition \(Q_\rho=9\), Eq.~\eqref{eq:fixed-Q}.

Under \(A_\mu\mapsto bA_\mu\), \(P[A]\) scales quadratically and \(S_{B,0}[A]\) quartically. Thus, for \(P[A]\neq0\),

\begin{equation}
\mathcal I[A]:=\frac{S_{B,0}[A]}{P[A]^2}
\label{eq:shape-action}
\end{equation}
is scale invariant, and stationarity at fixed \(P=q_0>0\) corresponds to stationarity of this shape action. This local fixed-scale construction is distinct from both the unconstrained Lorentzian path integral and proposals to nonperturbatively fix the noncompact Lorentz symmetry \cite{asano2025}. We remove Lorentz-orbit directions locally near the reference background when defining the one-loop measure.

A non-exhaustive numerical search for solutions of the fixed-scale classical equations at finite \(N=5\) identified at least 23 distinct stationary-solution classes. One-real-parameter local solution families were numerically constructed in four of these classes. Among these branches, we use the local family studied in Ref.~\cite{muramatsu2026} as a representative background: its regular local continuation, previously studied fluctuation properties, and explicit internal structure make it suitable for the present local analysis. This branch contains a configuration with \(A_0=0\) in an appropriate Lorentz rest frame and a pre-existing \(3+6\) spatial hierarchy.

Nonlinear numerical continuation of a Hessian null direction gives the selected family \(A_\mu(u)\), whose real modulus \(u\) cannot be removed by a color-gauge transformation or change of Lorentz frame. We take

\begin{equation}
A_\mu^{(0)}:=A_\mu(0)
\label{eq:reference-background}
\end{equation}
as the reference background.

We use \(A^{(0)}\) for the numerical reference configuration and \(u\) for the local coordinate retained in the one-loop density. The analytic classical modulus introduced below has a separate phase coordinate \(\theta_{\rm mod}\).

\subsection{Timelike kernel direction and spatial hierarchy of the reference background}\label{sec:background}

To examine the extent of the reference background in each spacetime direction, we introduce the quadratic matrix moments

\begin{equation}
G_{\mu\nu}[A]:=\frac1N\operatorname{Tr}(A_\mu A_\nu).
\label{eq:moments-G}
\end{equation}

The matrix \(G_{\mu\nu}\) is real symmetric and positive semidefinite. Using the Lorentzian metric, we define

\begin{equation}
K^\mu{}_{\nu}[A]:=\eta^{\mu\rho}G_{\rho\nu}[A].
\label{eq:moments}
\end{equation} The eigenvalues of \(K\) are independent of the Lorentz frame and characterize the temporal and spatial extents of the background.

For the selected reference background, numerical analysis verifies that \(G\) has rank 9 and that its one-dimensional kernel is timelike.
Writing this timelike kernel direction as a unit vector \(n^\mu\), we have \(n^\mu n_\mu=-1\) and \(K^\mu{}_{\nu}n^\nu=0\). It follows that \(G_{\mu\nu}n^\nu=0\).

In particular, since \(n^\mu A_\mu^{(0)}\) is Hermitian,

\begin{equation}
\operatorname{Tr}\left[(n^\mu A_\mu^{(0)})^2\right]=0
\label{eq:time-null-proof}
\end{equation} implies \(n^\mu A_\mu^{(0)}=0\).
Thus, \(A_0^{(0)}=0\) in a Lorentz rest frame whose time axis is aligned with \(n^\mu\). We use this frame below.

The spatial directions of the reference background already exhibit a hierarchy of three larger and six smaller directions.
Ordering the nine positive eigenvalues of \(K\) as \(\lambda_1^{K,+},\ldots,\lambda_9^{K,+}\) from largest to smallest, we define the spatial hierarchy ratio by

\begin{equation}
R_{3+6}^{K}
:=\frac{\frac13\sum_{i=1}^{3}\lambda_i^{K,+}}{\frac16\sum_{i=4}^{9}\lambda_i^{K,+}}.
\label{eq:spatial-ratio}
\end{equation} The scale-independent ratio compares the mean of the three largest eigenvalues with that of the remaining six. The label \(3+6\) denotes this ordered hierarchy, not exact degeneracy within either group.

We also define the eigenvalue gap at the boundary between the two groups by

\begin{equation}
\Delta_{3|4}^{K}:=\lambda_3^{K,+}-\lambda_4^{K,+}.
\label{eq:spatial-gap}
\end{equation}

At the reference background, \(R_{3+6}^{K}\simeq1.43\) and \(\Delta_{3|4}^{K}\simeq0.0990\). We study their response to \(A_0\) deformations below; residual and spectral diagnostics are summarized in Appendix~\ref{app:A-quotient}.

An exact constrained stationary solution near the numerical reference admits an analytic phase family along which the quadratic moments \(G_{\mu\nu}\) are constant in a fixed reflection-adapted frame. Consequently, the Lorentz-invariant spectrum of \(K\) and its associated hierarchy ratio and third--fourth gap remain constant along the family. For each of the 23 numerical configurations on the selected branch, a pointwise rigorous bound places it within \(1.65\times10^{-14}\) of this common analytic family in the positive Frobenius norm of all ten matrices, after individually determined \(SU(5)\) gauge and proper, orthochronous Lorentz transformations. This is a pointwise proximity result for samples of one family, distinct from the 23 solution classes of the broader search. Appendix~\ref{app:B-certificate} gives the enclosure and comparison bounds, and Appendices~\ref{app:B-phase} and~\ref{app:B-invariants} establish the phase family and its invariants; the 23 configurations are numerical approximations.

\subsection{Fifteen independent temporal matrix deformation directions}\label{sec:dimensions}

At the reference point \(A_0^{(0)}=0\), the traceless Hermitian matrix \(A_0\) has 24 real degrees of freedom. Infinitesimal Lorentz boosts change \(A_0\) along the span of the nine independent spatial matrices \(A_i^{(0)}\). Removing these frame directions leaves a 15-dimensional temporal matrix-deformation space \(\mathcal T_{\rm time}\). These are directions in matrix-configuration space, not additional spacetime directions.

We choose a basis \(H_a\) (\(a=1,\ldots,15\)) orthonormal with respect to the trace inner product and write the deformation of \(A_0\) as

\begin{equation}
\delta A_0=\sum_{a=1}^{15}c^aH_a.
\label{eq:time-deformation}
\end{equation}

Here \(c^a\) are real coefficients in a basis satisfying \(\operatorname{Tr}(H_aH_b)=\delta_{ab}\) and \(\operatorname{Tr}(H_aA_i^{(0)})=0\). This positive-definite trace inner product is auxiliary: it measures deformation magnitudes in matrix-configuration space and does not replace the Lorentzian spacetime metric.

The ten traceless Hermitian \(5\times5\) matrices have 240 real degrees of freedom. At the selected solution, the gauge-orbit tangent map has rank 24, the fixed-scale constraint removes one direction, and the Lorentz tangent map has rank 45 after gauge projection. In Ref.~\cite{muramatsu2026}, the Lorentz-orbit tangent rank was numerically checked at 23 points along this one-parameter solution family. These points belong to one family and should not be confused with the at least 23 distinct stationary-solution classes identified in the broader search.

The dimension of the physical fluctuation space is therefore

\begin{equation}
240-24-1-45=170.
\label{eq:dimension-chain}
\end{equation}

Separating the single modulus zero mode identified in the numerical Hessian leaves 169 nonzero-curvature directions. The temporal block is numerically nondegenerate at the reference background, so its 15 directions have a 154-dimensional Hessian-orthogonal complement. Numerical quotient diagnostics are given in Appendix~\ref{app:A-quotient}. Section~\ref{sec:temporal-extent} prescribes the temporal deformations and imposes classical stationarity in those 154 residual directions.

\section{Classical temporal extent under \texorpdfstring{\(A_0\)}{A0} deformations}\label{sec:temporal-extent}

Starting from the rest-frame background with \(A_0=0\), we prescribe deformations in the 15 independent temporal-matrix directions of Section~\ref{sec:dimensions} and determine the residual response by classical stationarity. We then examine the resulting timelike extent. The deformation parameter labels nearby matrix configurations; it is not an external time coordinate.

The temporal coefficients below are varied on the original section \(u=0\); the static analytic modulus is treated separately in Appendix~\ref{app:B}.

\subsection{Conditional classical response to \texorpdfstring{\(A_0\)}{A0} deformations}\label{sec:conditional}

We write the 15 deformation coefficients of the preceding section as \begin{equation}
c=rd,\qquad \|d\|_+=1.
\label{eq:direction-amplitude}
\end{equation}

Here \(|r|\) is the deformation amplitude measured with the trace inner product, and \(d\) is its unit direction. This normalization is defined in matrix-configuration space and is distinct from the spacetime Lorentzian metric.

At the reference background, these 15 directions are tangent to the fixed-scale surface, and the restricted Hessian \(H_{\rm time}\) is nondegenerate. They therefore embed in the 169-dimensional nondegenerate fluctuation space obtained by separating the physical modulus zero mode. Taking the 154-dimensional residual space to be their Hessian-orthogonal complement gives an invertible residual Hessian \(H_{\rm res}\) and a vanishing mixed block \(H_{\rm mix}\) (Appendix~\ref{app:A-splitting}). This orthogonality is distinct from the trace inner product used to normalize the amplitude.

Let \(u\) be the modulus coordinate and \(y=(y^1,\ldots,y^{154})\) the residual coordinates. On the local section \(u=0\), we prescribe the 15 coefficients \(c\) and solve the 154 residual stationarity equations for \(y(c)\); stationarity is not imposed in the prescribed temporal directions. This defines the conditional classical response used here.

For an exact constrained stationary base with this smooth chart and invertible residual Hessian, the implicit function theorem gives a locally unique \(y(c)\). At the numerical reference background, the invertibility of the 154-dimensional residual Hessian is supported by the computed inertia in Appendix~\ref{app:A-quotient}. With \(H_{\rm mix}=0\), the first derivative is

\begin{equation}
D_cy(0)=-H_{\rm res}^{-1}H_{\rm mix}=0
\end{equation}
and hence
\begin{equation}
y(c)=O(\|c\|_+^2).
\label{eq:quadratic-response}
\end{equation}

The prescribed deformation of \(A_0\) starts at first order, whereas the residual matrix components respond from second order. In the next subsection, we examine the temporal extent of this configuration through the eigenvalues of \(K\).

\subsection{Quadratic and quartic expansion of the timelike eigenvalue}\label{sec:temporal}

Let \(\lambda_t^K\) be the eigenvalue of \(K^\mu{}_{\nu}\) that continues from the timelike zero eigenvalue at the reference background. In our mostly-plus convention, \(\lambda_t^K\) is negative near the reference point, and \(-\lambda_t^K\) measures the magnitude of the temporal extent.

First, write the matrices before fixed-scale normalization as \(B_\mu\). Their temporal component along the 15 deformation directions is

\begin{equation}
B_0=rH(d),\qquad H(d):=\sum_{a=1}^{15}d^aH_a,
\end{equation} while the conditional classical response of the preceding subsection gives \(B_i=A_i^{(0)}+O(r^2)\) for the spatial components. The trace normalization of the basis and its orthogonality to the background spatial matrices imply

\begin{equation}
G_{00}[B]=\frac{r^2}{N},
\label{eq:moment-orders}
\end{equation} \begin{equation}
G_{0i}[B]=O(r^3),
\end{equation} \begin{equation}
G_{ij}[B]=G_{ij}[A^{(0)}]+O(r^2).
\end{equation} More explicitly, let \(G_{\rm sp}:=(G_{ij})_{i,j=1}^{9}\) be the spatial block and \(w_i:=G_{0i}[B]\) the mixing vector. The timelike eigenvalue satisfies

\begin{equation}
\lambda_t^K[B]
=-\frac{r^2}{N}
+w^{\mathsf T}\left(G_{\rm sp}[B]-\lambda_t^K[B]I_9\right)^{-1}w.
\end{equation} The spatial block at the reference background is positive and nondegenerate, so the inverse matrix is bounded in a neighborhood of the reference point. Since \(w=O(r^3)\), the eigenvalue before normalization obeys

\begin{equation}
\lambda_t^K[B]=-\frac{r^2}{N}+O(r^6).
\end{equation}

Next, we uniformly rescale the configuration to satisfy the fixed-scale condition \(Q_\rho=9\). Set \(q_{\rm raw}:=\operatorname{Tr}(B_\mu B^\mu)\) and \(A_\mu=\beta B_\mu\), with \(\beta^2=q_0/q_{\rm raw}\). The fixed-scale tangent condition for the residual columns is \(\sum_i\operatorname{Tr}(A_i^{(0)}E_{ip})=0\). Substituting \(B_i=A_i^{(0)}+\sum_p y^pE_{ip}\) therefore eliminates the cross term between the background and the residual response, while the square of \(y=O(r^2)\) starts at fourth order. Thus,

\begin{equation}
q_{\rm raw}=q_0-r^2+O(r^4)
\end{equation} and the normalization factor satisfies

\begin{equation}
\beta^2=1+\frac{r^2}{q_0}+O(r^4).
\label{eq:scale-expansion}
\end{equation} Here \(r\) is the deformation parameter before normalization; the normalized temporal component is \(A_0=\beta rH(d)\).

The eigenvalues of \(K\) are multiplied by \(\beta^2\) under this uniform matrix rescaling. The local expansion of the timelike eigenvalue at fixed \(Q_\rho=9\) is therefore

\begin{equation}
\lambda_t^K(r,d)
=-\frac{r^2}{N}-\frac{r^4}{Nq_0}
+\ell_6(d)r^6+O(r^8).
\label{eq:time-eigenvalue}
\end{equation} Here \(N=5\), \(q_0=15+2\sqrt3\), and \(\ell_6(d)\) is the sixth-order coefficient. Time-reflection covariance of the residual stationarity equations and uniqueness of the local solution imply \(y(-c)=y(c)\), so only even orders occur in the eigenvalue expansion (Appendix~\ref{app:C-reflection}).

The trace normalization fixes the quadratic coefficient, and fixed-scale normalization fixes the quartic coefficient. All 15 directions therefore give the same temporal extent through fourth order when compared at the same unnormalized amplitude \(r\); direction dependence can enter through \(\ell_6(d)\). This common classical behavior provides a reference for the directional response of the one-loop weight.

\begin{figure}[htbp]
\centering
\includegraphics[width=0.78\linewidth]{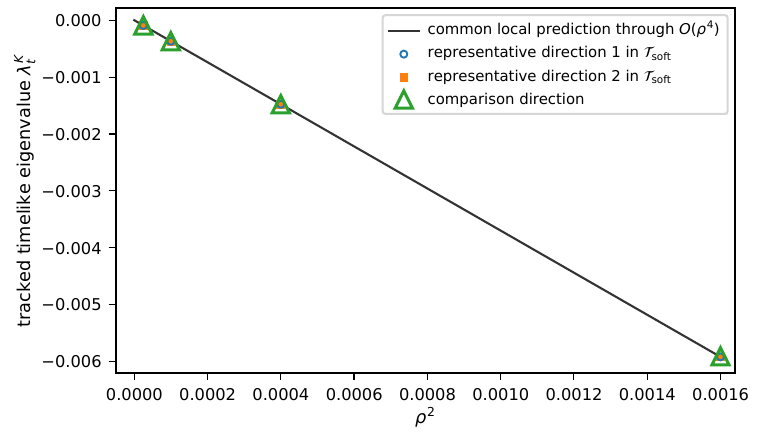}
\caption{Timelike eigenvalue $\lambda_t^K$ along two representative $A_0$ deformation directions in the soft temporal two-plane $\mathcal T_{\rm soft}$ and one comparison direction. The solid line shows the common analytical prediction through fourth order, and the markers are numerical results for conditional classical configurations at fixed $Q_\rho=9$. The two soft-plane directions are nearly coincident at the displayed scale.}
\label{fig:temporal-eigenvalue}
\end{figure}

In Figure~\ref{fig:temporal-eigenvalue}, the horizontal axis is the square of the dimensionless amplitude \(\rho:=r/\sqrt{q_0}\). Through fourth order, the analytical result above reads

\begin{equation}
\lambda_t^K=-\frac{q_0}{5}(\rho^2+\rho^4)+O(\rho^6).
\end{equation} Figure~\ref{fig:temporal-eigenvalue} compares the timelike eigenvalue along three representative deformation directions. The first two are the direction obtained by projecting the one-loop gradient (defined in Section~\ref{sec:nonpf}) onto the distinguished two-dimensional plane and the orthogonal direction within that plane. For comparison, we also consider a direction along which the spatial eigenvalue gap decreases at quadratic order (Section~\ref{sec:spatial}). Although the spatial responses differ among these directions, the timelike eigenvalue has the same behavior through fourth order.

\FloatBarrier

\section{Two-dimensional real plane determined by background geometry and fermionic structure}\label{sec:geometry}

Section~\ref{sec:temporal-extent} showed that temporal extent has direction-independent quadratic and quartic coefficients for equal-amplitude deformations in the 15-dimensional \(A_0\) space. We now ask whether the local one-loop weight can distinguish these directions at first order. We first define the retained-coordinate one-loop response, then construct a real two-plane from the background spatial matrices and a related fermionic subspace. Section~\ref{sec:mechanism} uses these structures to analyze the Pfaffian response.

\subsection{One-loop weight and first-order response in deformation directions}\label{sec:oneloop}

For each prescribed set of 15 coefficients \(c^a\), conditional classical stationarity determines the 154 residual coordinates and hence a configuration \(A_\mu(c)\) (Section~\ref{sec:conditional}). To compare local one-loop weights, we retain \(c\) and the classical modulus while integrating quadratic bosonic fluctuations over the 154 residual directions. Our previous paper instead integrated all 169 nonzero-curvature directions. Integrating the retained 15 directions at the reference point in the same quadratic approximation reproduces the leading local 169-direction Gaussian density and phase, including the coordinate measure (Appendix~\ref{app:C-gaussian}).

The Lorentzian quadratic integral is oscillatory. We define this finite-dimensional integral by adding an infinitesimal positive-definite damping factor and taking the zero-damping limit (Appendix~\ref{app:C-fresnel}).

The fermionic contribution is obtained from the Grassmann integral over the Majorana--Weyl fermions introduced in Section~\ref{sec:model-scale}. For a matrix configuration \(A\), define

\begin{equation}
\mathcal M_F[A]:=\widehat\gamma^\mu\otimes D_\mu^A.
\end{equation} Here \(\widehat\gamma^\mu=i\gamma^\mu\) are 16-dimensional real symmetric matrices representing the bilinear of ten-dimensional Majorana--Weyl spinors. We use the convention \(\widehat\gamma^0=I_{16}\) for the temporal component and \(\{\widehat\gamma^i,\widehat\gamma^j\}=2\delta^{ij}I_{16}\) for the spatial components. The derivative \(D_\mu^AX=i[A_\mu,X]\) acts on the 24-dimensional adjoint matrix space. The matrix \(\mathcal M_F[A]\) is real antisymmetric and 384-dimensional, and the configuration-dependent factor produced by the fermionic integral is \(\operatorname{Pf}\mathcal M_F[A]\). These basis and Pfaffian-sign conventions are held fixed throughout. As shown in our previous paper \cite{muramatsu2026}, the magnitude of the Pfaffian for a complete matrix configuration is invariant under color-gauge transformations and proper Lorentz transformations connected to the identity.

The resulting local one-loop density is defined with respect to \(d^{15}c\,du\); the 15 deformation coefficients and the classical modulus are retained rather than integrated. Here \(u\) is the original retained coordinate: introducing \(\theta_{\rm mod}\) in the classical construction does not change this density chart or insert a Jacobian. Fixing \(u=0\), we denote by \(L_{\rm1loop}(c)\) the logarithm of the magnitude of the combined fermionic Pfaffian, residual bosonic fluctuation, gauge-fixing, and local-measure factors, omitting deformation-independent constants. Here ``full one-loop'' means that all these local factors are included, not that the retained coordinates have been integrated. The Lorentzian classical-action and Gaussian phases are distinct from the magnitude gradient studied here. Around \(c=0\),

\begin{equation}
L_{\rm1loop}(c)-L_{\rm1loop}(0)
=\sum_{a=1}^{15}J_{{\rm1L},a}c^a+O(\|c\|_+^2).
\end{equation} The coefficient \(J_{{\rm1L},a}\) is the first-order change under an infinitesimal variation of \(c^a\) with the other \(c^b\) held fixed, defined by

\begin{equation}
J_{{\rm1L},a}:=
\left.\frac{\partial L_{\rm1loop}(c)}{\partial c^a}\right|_{c=0}.
\end{equation} Applying the same operation to the Pfaffian factor alone defines

\begin{equation}
J_{{\rm Pf},a}:=
\left.\frac{\partial}{\partial c^a}
\log\bigl|\operatorname{Pf}\mathcal M_F[A(c)]\bigr|\right|_{c=0}.
\end{equation} The partial derivative is taken along \(A(c)\), as determined by the conditional classical stationarity equations. In a local region where the Pfaffian is nonzero, the identity between its square and the determinant gives

\begin{equation}
J_{{\rm Pf},a}=
\frac12\operatorname{Re}\operatorname{Tr}\!\left[
\mathcal M_F[A^{(0)}]^{-1}
\left.\frac{\partial\mathcal M_F[A(c)]}{\partial c^a}\right|_{c=0}
\right].
\label{eq:local-response}
\end{equation} We collect these sets of 15 coefficients into \(J_{\rm1L}\) and \(J_{\rm Pf}\), respectively, and compare their directions and magnitudes using the auxiliary positive-definite inner product associated with the trace-orthonormal basis of Section~\ref{sec:dimensions}. The vector \(J_{\rm1L}\) describes the first-order response of the logarithm of the magnitude of the total one-loop weight, while \(J_{\rm Pf}\) describes the response of its Pfaffian factor. Defining them separately allows us to compare the directional sensitivity of the total local one-loop weight with that of its Pfaffian factor under \(A_0\) deformations. To relate the Pfaffian variation to fermionic eigenvalues and eigenstates, we first examine the structure inherited by the fermionic operator from the background geometry.

To identify the fermionic states responsible for the Pfaffian variation, we use the Hermitian form of the operator at the reference configuration,

\begin{equation}
\mathcal K_F:=i\mathcal M_F[A^{(0)}].
\end{equation} The operator \(\mathcal K_F\) is 384-dimensional, with eigenvalues in positive--negative pairs. We focus on the lower end of the positive spectrum of \(\mathcal K_F\), in particular on the structure of the lowest positive-eigenvalue doublet and its eigenvalue sensitivity to temporal matrix deformations. Since \(D_cy(0)=0\), \(D_c\beta(0)=0\), and \(\widehat\gamma^0=I_{16}\), the first-order variation at the reference point under \(\delta A_0=H\) is \(\delta\mathcal K_F=iI_{16}\otimes D_H\) (Appendix~\ref{app:D-sensitivity}). Before analyzing it, we construct from the background spatial matrices the two-dimensional real plane that will serve as a reference for comparison.

\subsection{Two-dimensional real plane determined by the internal background structure}\label{sec:plane}

The 15 deformation directions of \(A_0\) need not couple in the same way to the background spatial matrices. We therefore examine the common color-space structure of the nine spatial matrices at the reference configuration.

The numerically identified color reflection splits the five-dimensional color space into one- and four-dimensional subspaces and separates the nine spatial matrices into three odd and six even matrices. The original root search did not impose this reflection; the nearby exact solution realizes it by the construction in Appendix~\ref{app:B-existence}. The following algebra applies under this exact block structure, which is satisfied by the numerical reference background to the accuracy of the calculation. In a reflection-adapted basis, the odd matrices connect the two color subspaces and the even matrices act within them:

\begin{equation}
A_a^{(0)}=
\begin{pmatrix}0&z_a^\dagger\\z_a&0\end{pmatrix}
\quad(a=1,2,3)
\end{equation} and

\begin{equation}
A_b^{(0)}=
\begin{pmatrix}a_b&0\\0&B_b\end{pmatrix}
\quad(b=4,\ldots,9).
\end{equation} Here \(z_a\) is a four-component complex column vector, \(a_b\) is real, and \(B_b\) is a four-dimensional Hermitian matrix. This three--six classification is based on the color-space reflection and differs from the eigenvalue classification into three larger and six smaller spacetime directions introduced in Section~\ref{sec:background}. Numerical analysis of the selected background shows that the three directions of larger extent lie in the sector of the six reflection-even matrices.

We now apply this color-space decomposition to deformations of \(A_0\). Write a Hermitian matrix connecting the one- and four-dimensional components as

\begin{equation}
c(w):=\begin{pmatrix}0&w^\dagger\\w&0\end{pmatrix},
\qquad w\in\mathbb C^4.
\end{equation} It has eight real degrees of freedom, corresponding to the real and imaginary parts of \(w\). Among them, the three real directions along the reflection-odd background matrices are changes of frame generated by Lorentz boosts. With \(\mathsf Z:=(z_1,z_2,z_3)\), the trace-orthogonality condition removing these boosts is

\begin{equation}
\operatorname{Re}(\mathsf Z^\dagger w)=0.
\end{equation} Since \(\mathsf Z\) has complex rank 3, these three real conditions leave a five-real-dimensional subspace of \(A_0\) deformations. Trace orthogonality to the six reflection-even background matrices follows from the block structure.

Within this five-real-dimensional subspace, we seek directions on which the double commutator with the background acts particularly simply. Define the contribution of the three reflection-odd matrices by

\begin{equation}
\Delta_3X:=\sum_{a=1}^{3}[A_a^{(0)},[A_a^{(0)},X]].
\end{equation} Its action on \(c(w)\) contains terms depending on \(\mathsf Z^\dagger w\). The boost-removal condition above imposes three real constraints; the following stronger complex-orthogonality condition identifies a smaller subspace on which those terms vanish:

\begin{equation}
\mathsf Z^\dagger w=0.
\end{equation} Under this condition,

\begin{equation}
\Delta_3c(w)=\tau_3c(w),
\qquad \tau_3:=\operatorname{Tr}(\mathsf Z^\dagger\mathsf Z),
\end{equation} so the action of the three odd background matrices reduces to scalar multiplication in these directions. Since \(\mathsf Z\) has complex rank 3, the kernel in \(\mathbb C^4\) is one-complex-dimensional. For a normalized kernel vector \(z_0\), the matrices \(c(z_0)\) and \(c(iz_0)\) are two independent real matrix deformations. We define the plane they span by

\begin{equation}
P_0:=\{c(w):\mathsf Z^\dagger w=0\},
\qquad \dim_{\mathbb R}P_0=2.
\label{eq:kernel-plane}
\end{equation} This is a two-dimensional real plane within the 15-dimensional matrix-deformation space. A phase change of \(z_0\) merely rotates the two basis vectors, leaving the plane unchanged. Within the five-real-dimensional reflection-odd physical deformation space, this plane is also characterized as the kernel of the positive-semidefinite correction to the scalar part of \(\Delta_3\).

The three odd matrices thus act simply on this plane. We next ask whether the same plane remains invariant when the six even matrices are included. Acting on \(c(w)\), the double commutator with a block-diagonal \(A_b^{(0)}\) produces \((B_b-a_bI_4)^2w\). We collect these six contributions into

\begin{equation}
K_6:=\sum_{b=4}^{9}(B_b-a_bI_4)^2.
\end{equation} The classical stationarity equations of the reference background imply that \(K_6\) preserves the complex subspace spanned by the three columns of \(\mathsf Z\). Since \(K_6\) is Hermitian, it also preserves their orthogonal complement, the one-dimensional complex kernel, and hence

\begin{equation}
K_6z_0=\kappa_6z_0,
\qquad \kappa_6\in\mathbb R.
\end{equation} Thus, the double commutators with the six matrices also act as a scalar on \(P_0\). The algebraic derivations of the plane characterization and this stationary invariance are given in Appendix~\ref{app:A-plane}.

In the trace-orthonormal deformation coefficients, the constrained Hessian on the 15-dimensional \(A_0\) deformation space is the restriction of the quadratic form \(H_{\rm time}=2\lambda_QI-\Delta\), where \(\Delta X:=\sum_{i=1}^{9}[A_i^{(0)},[A_i^{(0)},X]]\). Since both the three-matrix and six-matrix contributions are scalar on \(P_0\), we obtain

\begin{equation}
H_{\rm time}\big|_{P_0}=h_0I_2
\label{eq:plane-curvature}
\end{equation} and

\begin{equation}
h_0=2\lambda_Q-\tau_3-\kappa_6.
\end{equation} The construction of the plane and the scalar action of the Hessian are algebraic results based on the background block structure, complex rank 3, and the stationarity conditions.

At the selected reference background, numerical evaluation gives \(h_0\simeq0.0432>0\) and identifies \(P_0\) with the lowest positive-curvature doublet of the 15-dimensional temporal Hessian. Equality of the two curvatures follows algebraically; their positivity and ordering follow numerically. We also write \(\mathcal T_{\rm soft}:=P_0\). Here ``soft'' means small positive curvature at the selected background, not a Hessian zero mode.

Along the exact phase family, the plane \(P_0\) and its scalar curvature \(h_0\) remain fixed (Appendix~\ref{app:B-invariants}). These are invariants in matrix-configuration space; the positivity and ordering above are numerical properties of the reference background.

\subsection{Fermionic subspace associated with the same background color direction}\label{sec:fermion}

The two-plane is determined by a complex color direction \(z_0\) orthogonal to the three reflection-odd background columns. To relate it to the fermionic operator, which acts on spinor and adjoint color components, we first form a traceless color matrix from \(z_0\).

Normalize \(z_0^\dagger z_0=1\) and embed it in the five-dimensional color space as \(\widetilde z_0:=(0,z_0)^{\mathsf T}\). The projector onto this one-dimensional complex direction is

\begin{equation}
F_z:=\widetilde z_0\widetilde z_0^\dagger.
\end{equation} We then define a traceless Hermitian matrix of unit norm with respect to the trace inner product by

\begin{equation}
H_z:=\sqrt{\frac54}\left(F_z-\frac15I_5\right).
\end{equation} The matrix \(H_z\) represents the background-determined color direction as a single matrix in the adjoint representation.

To distinguish the coupling of this color direction to the other color components, we use

\begin{equation}
W_z:=I_5-2F_z.
\end{equation} The reflection \(W_z\) multiplies the direction \(\widetilde z_0\) by \(-1\) and its orthogonal complement by \(+1\). Whereas \(R\) in Section~\ref{sec:plane} splits the full color space as \(1+4\), \(W_z\) singles out a particular one-dimensional direction inside that four-dimensional sector. The two reflections distinguish different color subspaces.

Decompose each background spatial matrix into parts even and odd under \(W_z\). The even part is

\begin{equation}
A_i^+:=\frac12(A_i^{(0)}+W_zA_i^{(0)}W_z)
\end{equation} and is obtained by removing the matrix components connecting \(\widetilde z_0\) to its orthogonal complement. Thus, \(A_i^+\) commutes with \(F_z\). Since \(H_z\) is linear in \(F_z\),

\begin{equation}
D_i^+H_z=i[A_i^+,H_z]=0,
\end{equation} where \(D_i^+X:=i[A_i^+,X]\).

We apply this property to the fermionic operator. In addition to the original operator \(\mathcal K_F\), define an auxiliary operator constructed only from \(A_i^+\) by

\begin{equation}
\mathcal K_{F,+}:=i\widehat\gamma^i\otimes D_i^+.
\end{equation} Taking the color component to be \(H_z\) for any of the 16 spinor components gives \(D_i^+H_z=0\), so \(\mathcal K_{F,+}\) annihilates the resulting state. For the spectral analysis, denote the complexification of the 16 real Majorana--Weyl spinor coefficients by \(S_+\cong\mathbb C^{16}\). Then

\begin{equation}
\mathcal Z_F:=S_+\otimes\operatorname{span}_{\mathbb C}\{H_z\},
\qquad \dim_{\mathbb C}\mathcal Z_F=16
\end{equation} satisfies

\begin{equation}
\mathcal K_{F,+}\mathcal Z_F=0.
\label{eq:auxiliary-null}
\end{equation} This is a zero-mode subspace of the auxiliary operator constructed from the even components. It is distinct from the timelike kernel of the bosonic moment matrix \(G\) and from the physical-modulus zero mode of the bosonic Hessian. The color direction \(z_0\) entering this auxiliary subspace is the same one that independently defines \(P_0\) in Section~\ref{sec:plane}, relating the classical matrix deformations to the fermionic structure.

The original 384-dimensional fermionic operator \(\mathcal K_F\) also contains the matrix components odd under \(W_z\). Consequently, components in \(\mathcal Z_F\) couple to its 368-dimensional orthogonal complement. To display this coupling, write the original operator in a basis adapted to \(\mathcal Z_F\) and its orthogonal complement as

\begin{equation}
\mathcal K_F\sim
\begin{pmatrix}0&B_F^\dagger\\B_F&C_F\end{pmatrix}.
\label{eq:full-fermion-block}
\end{equation} Here \(B_F\) describes the coupling between the two subspaces, and \(C_F\) is the restriction to the 368-dimensional complement. The zero diagonal block follows because the auxiliary operator annihilates \(\mathcal Z_F\) and the odd components connect the two subspaces.

In the eigenvalue equation \(\mathcal K_F(x,y)^{\mathsf T}=\lambda(x,y)^{\mathsf T}\), when \(C_F-\lambda\) is invertible, the complementary component is determined by

\begin{equation}
y=-(C_F-\lambda)^{-1}B_Fx.
\end{equation} Substituting this into the remaining equation gives

\begin{equation}
-B_F^\dagger(C_F-\lambda)^{-1}B_Fx=\lambda x.
\end{equation} This spectral-parameter-dependent equation is a nonlinear spectral condition on the 16-dimensional auxiliary subspace: the resolvent propagates its coupling through the 368-dimensional complement and back. It retains the effect of that complement rather than replacing the full operator by a fixed 16-dimensional matrix.

Numerical diagonalization of the 384-dimensional fermionic operator at the reference background gives a doublet at the lowest positive eigenvalue \(\lambda_*\simeq0.0682\). Let \(P_*\) project onto the entire lowest doublet and \(P_Z\) onto the 16-dimensional \(\mathcal Z_F\). The doublet-averaged squared-norm fraction is

\begin{equation}
w_Z:=\frac12\operatorname{Tr}(P_*P_Z)\simeq0.2995.
\end{equation}

The denominator 2 is the multiplicity of the lowest positive doublet, not the dimension of \(\mathcal Z_F\). Thus approximately 30\% of the doublet-averaged squared norm lies in the auxiliary subspace, and 70\% lies in its 368-dimensional complement. The doublet must therefore be analyzed using eigenstates of the full fermionic operator.

\section{From fermionic eigenvalue sensitivity to the full one-loop directional response}\label{sec:mechanism}

The two-plane \(P_0\) and auxiliary fermionic subspace \(\mathcal Z_F\) share the background color direction \(z_0\). We now examine the first-order sensitivity of the full lowest positive doublet to temporal deformations and determine how this response enters the Pfaffian and the complete local one-loop weight. The spectral identifications, concentration fractions, and Pfaffian responses in this section are evaluated numerically at the reference background.

\subsection{First-order sensitivity of the lowest positive-eigenvalue doublet}\label{sec:sensitivity}

To determine which temporal directions most strongly change the lowest positive eigenvalues, define the Hermitian operator on the deformed configuration

\begin{equation}
\mathcal K_F(c):=i\mathcal M_F[A(c)].
\end{equation} At the reference point \(c=0\), let \(P_r\) be the projector onto an eigenspace with positive eigenvalue \(\lambda_r\) and multiplicity \(d_r\). The eigenvalues may split under deformation, but the first derivative of their mean \(\bar\lambda_r(c)\) is fixed by the full reference-point projector:

\begin{equation}
a_{r,j}:=\left.\frac{\partial\bar\lambda_r(c)}{\partial c^j}\right|_{c=0}
\end{equation} is given by

\begin{equation}
a_{r,j}=\frac1{d_r}\operatorname{Tr}\!\left[
P_r\left.\frac{\partial\mathcal K_F(c)}{\partial c^j}\right|_{c=0}
\right],\qquad j=1,\ldots,15.
\label{eq:eigenvalue-sensitivity}
\end{equation} The complete projector makes this mean sensitivity independent of the basis chosen within the eigenspace. The vector \(a_r\) collects the sensitivities to the 15 temporal directions, compared with the trace inner product of Section~\ref{sec:dimensions}.

Let \(\Pi_0\) be the orthogonal projector onto the two-dimensional real plane \(P_0\) constructed from the background. The numerically computed sensitivity \(a_*\) of the lowest positive doublet of the 384-dimensional operator at the reference background satisfies

\begin{equation}
\frac{\|\Pi_0a_*\|^2}{\|a_*\|^2}\simeq0.9943\qquad(99.43\%).
\label{eq:sensitivity-fraction}
\end{equation} Thus, 99.43\% of the squared norm of the 15-dimensional sensitivity vector lies in the plane constructed independently from the spatial background. This is a numerical correspondence between background geometry and the full fermionic operator; the doublet classification is numerical at the resolved spectral precision (Appendix~\ref{app:D-numerics}).

\subsection{Internal structure of fermionic states and eigenvalue sensitivity}\label{sec:source}

The color and spin components of the lowest doublet enter its directional sensitivity through the full projector \(P_*\). A color--spin coupling vector relates this internal structure to the spatial background and its temporal Hessian.

To make the commutator relations with the background explicit, let \(e_\ell\) be 24 trace-orthonormal color matrices and define the operator associated with a variation in each color direction by

\begin{equation}
\mathcal J_\ell:=iI_{16}\otimes D_{e_\ell}.
\end{equation} Its double commutator with the fermionic operator separates into a color Laplacian built from the background spatial matrices and terms containing spin-matrix two-forms:

\begin{equation}
[\mathcal K_F,[\mathcal K_F,\mathcal J_\ell]]
=\mathcal J_{\Delta e_\ell}+\mathcal R_\ell,
\qquad
\Delta:=-\sum_{i=1}^{9}D_i^2.
\label{eq:color-spin-double-commutator}
\end{equation} Here \(\mathcal R_\ell\) is the Hermitian operator collecting the terms with distinct spatial spin indices. Its explicit definition and sign are given in Appendix~\ref{app:D-source}. Contracting this term with the eigenvalue-cluster projector defines the 24-component quantity

\begin{equation}
s_{r,\ell}:=-\frac1{d_r}\operatorname{Tr}(P_r\mathcal R_\ell).
\label{eq:source-definition}
\end{equation} We call \(s_r\) the color--spin coupling vector: it projects the spin-dependent commutator terms onto an eigenvalue cluster. The isometry \(\mathcal E\), whose columns form the trace-orthonormal basis of the 15 physical \(A_0\) deformation directions, restricts it to that space:

\begin{equation}
s_{r,\mathcal E}:=\mathcal E^{\mathsf T}s_r.
\end{equation}

Numerical evaluation for the lowest positive doublet also shows strong concentration of this 15-component color--spin coupling vector in the background two-dimensional real plane:

\begin{equation}
\frac{\|\Pi_0s_{*,\mathcal E}\|^2}
{\|s_{*,\mathcal E}\|^2}\simeq0.9835\qquad(98.35\%).
\label{eq:source-fraction}
\end{equation} This fraction is a different quantity from the 99.43\% obtained for the eigenvalue sensitivity in Section~\ref{sec:sensitivity}. The strong concentration already present at the level of the color--spin coupling vector shows that the combination of color and spin components in the lowest doublet is closely related to the characteristic deformation directions of this background.

We now connect this color--spin coupling vector to the eigenvalue sensitivity. Define the cluster-averaged trace of the operator \(\mathcal J_\ell\) for each color direction by

\begin{equation}
q_{r,\ell}:=\frac1{d_r}\operatorname{Tr}(P_r\mathcal J_\ell).
\end{equation} Since \(P_r\) is a spectral projector of \(\mathcal K_F\), it commutes with \(\mathcal K_F\). The projected trace of the double commutator therefore vanishes, and the operator identity above gives

\begin{equation}
\Delta q_r=s_r.
\label{eq:source-trace}
\end{equation}

On the other hand, the fixed-scale stationarity equations of the reference background imply that \(\Delta\) has eigenvalue \(2\lambda_Q\) on the color subspace spanned by the nine spatial matrices. On the 15-dimensional deformation space orthogonal to this nine-dimensional sector, \(\Delta\) is related to the constrained Hessian introduced in Section~\ref{sec:plane} by

\begin{equation}
\mathcal E^{\mathsf T}\Delta\mathcal E
=2\lambda_QI_{15}-H_{\rm time}.
\end{equation} The definition of the eigenvalue sensitivity also gives \(a_r=\mathcal E^{\mathsf T}q_r\). Combining these relations yields

\begin{equation}
(2\lambda_QI_{15}-H_{\rm time})a_r=s_{r,\mathcal E}.
\label{eq:source-sensitivity}
\end{equation} At the selected reference background, the numerically established full-rank gauge-orbit condition ensures this operator is invertible, so

\begin{equation}
a_r=(2\lambda_QI_{15}-H_{\rm time})^{-1}s_{r,\mathcal E}.
\end{equation}

The temporal Hessian thus converts the color--spin coupling vector into eigenvalue sensitivity. For the lowest doublet, their respective in-plane squared-norm fractions are 98.35\% and 99.43\%. An antiunitary reflection symmetry preserving complete spectral projectors restricts the projected color--spin coupling vector to a five-dimensional odd sector of the 15-dimensional temporal-deformation space. Appendix~\ref{app:D-sensitivity} gives the projection and invertibility conditions and the complete-projector selection rule. This analytic \(15\to5\) selection and the further numerical \(5\to2\) concentration are distinct results.

\subsection{Lowest positive doublet and first-order Pfaffian response}\label{sec:pfaffian}

Each exact positive eigenspace of eigenvalue \(\lambda_r\) and multiplicity \(d_r\) contributes \(d_ra_r/\lambda_r\) to the logarithmic Pfaffian-magnitude gradient. Summing over distinct positive eigenspaces gives

\begin{equation}
J_{\rm Pf}=\sum_r\frac{d_r}{\lambda_r}a_r.
\label{eq:pfaffian-sum}
\end{equation} For finite-precision spectral groups, contributions are given by an inverse-operator trace over each group projector rather than by a common-eigenvalue approximation (Appendix~\ref{app:D-numerics}). This gradient measures how the logarithm of the Pfaffian magnitude differs among matrix deformations of \(A_0\) of equal amplitude. As in Section~\ref{sec:temporal-extent}, set \(c=rd\), with \(\|d\|_+=1\). Near the reference point,

\begin{equation}
\log|\operatorname{Pf}\mathcal M_F[A(rd)]|
-\log|\operatorname{Pf}\mathcal M_F[A(0)]|
=r\,J_{\rm Pf}\cdot d+O(r^2).
\end{equation} At fixed \(r>0\), the first-order change is determined by the inner product of \(J_{\rm Pf}\) and \(d\): it is maximal in the parallel direction, zero in an orthogonal direction, and minimal in the antiparallel direction. The numerically obtained nonzero \(J_{\rm Pf}\) therefore directly shows that the Pfaffian distinguishes deformation directions of \(A_0\) already at first order.

Smaller positive eigenvalues carry larger inverse-eigenvalue weights. However, each term is a vector with a direction, so their sum must be evaluated to determine the total contribution.

The lowest positive eigenvalue \(\lambda_*\simeq0.0682\) forms a doublet, whose contribution is

\begin{equation}
J_*:=\frac{2}{\lambda_*}a_*.
\end{equation} Evaluating all 89 positive-eigenvalue clusters of the reference background gives the normalized signed projection of \(J_*\) onto the full Pfaffian gradient as

\begin{equation}
\frac{J_*\cdot J_{\rm Pf}}{\|J_{\rm Pf}\|^2}
\simeq0.9862\qquad(98.62\%).
\label{eq:signed-projection}
\end{equation} The relative vector residual when the full gradient is approximated by the lowest-doublet contribution alone is

\begin{equation}
\frac{\|J_{\rm Pf}-J_*\|}{\|J_{\rm Pf}\|}
\simeq0.0138\qquad(1.38\%).
\label{eq:vector-residual}
\end{equation} The signed projection measures alignment with the full gradient, whereas the residual measures the remaining vector contribution. They are distinct vector diagnostics, not complementary fractions of the Pfaffian. The lowest-doublet term thus approximates the full gradient with a 1.38\% relative vector residual.

\begin{figure}[htbp]
\centering
\begin{subfigure}{0.48\linewidth}
\centering
\includegraphics[width=\linewidth]{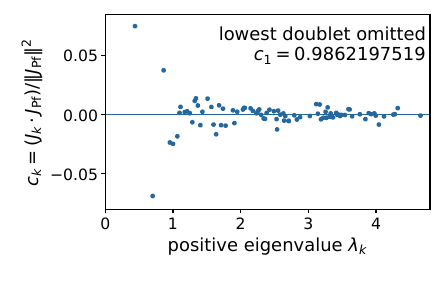}
\caption{Signed projection \(c_k:=(J_k\cdot J_{\rm Pf})/\|J_{\rm Pf}\|^2\) of the \(k\)-th spectral-group contribution \(J_k\). The lowest positive doublet, with \(c_1\simeq0.9862\), lies outside the displayed vertical range.}
\label{fig:pfaffian-panel-a}
\end{subfigure}\hfill
\begin{subfigure}{0.48\linewidth}
\centering
\includegraphics[width=\linewidth]{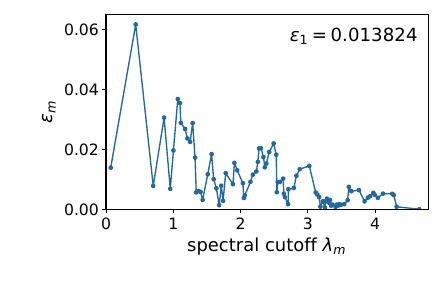}
\caption{Cumulative relative residual \(\varepsilon_m:=\|J_{\rm Pf}-\sum_{k=1}^{m}J_k\|/\|J_{\rm Pf}\|\) as the first \(m\) spectral groups are added in order of increasing positive eigenvalue.}
\label{fig:pfaffian-panel-b}
\end{subfigure}
\caption{Contributions of positive-eigenvalue clusters to the Pfaffian gradient.}
\label{fig:pfaffian-spectral-decomposition}
\end{figure}

As Figure~\ref{fig:pfaffian-spectral-decomposition} shows, the clusters above the lowest doublet have contributions of both signs, and the cumulative residual need not decrease monotonically. The spectral sum over all 89 numerical clusters agrees with a direct inverse-matrix-trace evaluation of the Pfaffian gradient to a relative difference of approximately \(1.5\times10^{-14}\). Details of the clustering and numerical consistency check are given in Appendix~\ref{app:D-numerics}.

\subsection{Full one-loop weight and first-order Pfaffian gradient}\label{sec:nonpf}

We now test whether the residual bosonic, gauge-fixing, and local-measure factors modify the first-order Pfaffian response. Decompose the logarithmic magnitude of the complete local one-loop density defined in Section~\ref{sec:oneloop} as

\begin{equation}
L_{\rm1loop}(c)=\log|\operatorname{Pf}\mathcal M_F[A(c)]|+L_{\rm nonPf}(c).
\end{equation} Here \(L_{\rm nonPf}\) contains the residual 154-direction Lorentzian Gaussian determinant, gauge fixing, and the local measure.

Since \(A_0^{(0)}=0\), reflection of the Lorentz time coordinate preserves the reference background and maps the 15 prescribed deformation coefficients of \(A_0\) as \(c\mapsto-c\). We treat the 154 residual directions and the retained deformation directions in the same reflection-adapted local coordinates. We also include in the local measure the gauge-transformation directions, Lorentz-transformation directions, fixed-scale normal direction, and classical modulus direction.

In this convention, let \(H_c\) be the constrained Hessian in the residual coordinates, \(F_A\) the frame matrix with all these coordinate directions as columns, and \(q_{\rm fix}=q_0\) the fixed-scale value. Combining the Faddeev--Popov factor with the corresponding inverse factor in the local measure gives, up to deformation-independent constants,

\begin{equation}
L_{\rm nonPf}(c)
=-\frac12\log|\det H_c(c)|
+\log|\det F_A(c)|
-\log(2q_{\rm fix}).
\label{eq:nonpf-density}
\end{equation} The individual factors and construction of the full coordinate frame are given in Appendix~\ref{app:C-measure}.

When the residual and retained spaces are reflection-covariant and the Hessian, Gram matrices, and coordinate frame are nonsingular, geometric time reflection makes the contributions relevant to the first-order non-Pfaffian variation even under \(c\mapsto-c\). Consequently,

\begin{equation}
\left.\frac{\partial L_{\rm nonPf}}{\partial c^a}\right|_{c=0}=0,
\qquad a=1,\ldots,15.
\end{equation} The relation between time reflection and the coordinate frame, and the conditions needed for this cancellation, are given in Appendix~\ref{app:C-reflection}. The time reflection used here is a geometric transformation of bosonic configurations, distinct from the reflection transformation of a single-chirality Pfaffian.

Since the full one-loop gradient is \(J_{\rm1L}=J_{\rm Pf}+\left.\nabla_cL_{\rm nonPf}\right|_0\), at the selected reference background we obtain

\begin{equation}
J_{\rm1L}=J_{\rm Pf}.
\label{eq:nonpf-cancellation}
\end{equation} This is an equality of first derivatives of logarithmic magnitudes at the reference point in all 15 temporal directions. Individual non-Pfaffian terms need not vanish, and their combined first-order cancellation does not remove higher-order dependence. This equality is formulated in the retained \((c,u)\) chart under reflection-covariance and local nonsingularity conditions, supported numerically in the full fluctuation space at the reference background. The rigorous existence proof in Appendix~\ref{app:B} concerns the static reflection sector; the analytic phase \(\theta_{\rm mod}\) remains distinct from the retained coordinate \(u\).

Using all 89 positive-eigenvalue clusters, the in-plane squared-norm fraction of the complete local one-loop gradient is

\begin{equation}
\frac{\|\Pi_0J_{\rm1L}\|^2}{\|J_{\rm1L}\|^2}
\simeq0.9941\qquad(99.41\%).
\label{eq:full-gradient-fraction}
\end{equation} This 99.41\% measures the in-plane concentration of the full one-loop gradient and is distinct from the 99.43\% found for the eigenvalue sensitivity of the lowest doublet.

The first-order variation also directly shows which orientations within the two-dimensional real plane are distinguished by the Pfaffian gradient. Choose an orthonormal basis \(e_1,e_2\) of the plane and write a unit direction as \(d(\varphi)=e_1\cos\varphi+e_2\sin\varphi\). If \(\varphi_*\) is the angle of the projected gradient \(\Pi_0J_{\rm Pf}\), then

\begin{equation}
J_{\rm Pf}\cdot d(\varphi)
=\|\Pi_0J_{\rm Pf}\|_+\cos(\varphi-\varphi_*).
\end{equation} Thus, even for equal-amplitude deformations of \(A_0\) within the plane, the first-order change in the logarithm of the Pfaffian magnitude varies with angle and is maximal at \(\varphi=\varphi_*\). This distinction between directions concerns the local first-order sensitivity at the reference point. The angle \(\varphi\) rotates a temporal deformation within \(P_0\); it is not the static classical phase \(\theta_{\rm mod}\).

\section{Geometric consequences: changes in the pre-existing \texorpdfstring{\(3+6\)}{3+6} spatial hierarchy}\label{sec:spatial}

We use the local one-loop gradient to prescribe a temporal-deformation direction and then examine the conditional classical spatial response along it. Let \(\Pi_0\) project onto the background two-plane \(\mathcal T_{\rm soft}\). The unit direction maximizing the first-order increase of \(L_{\rm1loop}\) within this plane is

\begin{equation}
\hat e_{\rm1L}:=\frac{\Pi_0J_{\rm1L}}{\|\Pi_0J_{\rm1L}\|_+}.
\end{equation} The denominator is the norm induced by the trace inner product on the matrix-deformation coefficient space introduced in Section~\ref{sec:dimensions}. Prescribing this direction, we set

\begin{equation}
c=r\hat e_{\rm1L},\qquad \rho:=\frac{r}{\sqrt{q_0}}
\end{equation} and impose classical stationarity in the 154 residual directions, retaining the prescribed \(A_0\) coefficients and rescaling to \(Q_\rho=9\) as in Section~\ref{sec:temporal}. The one-loop gradient specifies an externally prescribed direction at the original \(u=0\); the residual response is determined by classical stationarity. The coefficients below are evaluated at this reference point, separately from the static phase-family invariants.

We quantify the pre-existing spatial hierarchy using the nine ordered positive \(K\)-eigenvalues (Section~\ref{sec:background}). The ratio of the mean of the largest three to that of the remaining six is \(R_{3+6}^{K}\); the third--fourth eigenvalue gap is \(\Delta_{3|4}^{K}\). At the reference point, \(R_0\simeq1.43\) and \(\Delta_0\simeq0.0990\). Their local expansions are

\begin{equation}
R_{3+6}^{K}(\rho,d)=R_0+R_2(d)\rho^2+O(\rho^4)
\label{eq:spatial-expansion}
\end{equation} and

\begin{equation}
\Delta_{3|4}^{K}(\rho,d)=\Delta_0+g_2(d)\rho^2+O(\rho^4).
\end{equation} At the reference background, the fourth and fifth positive spatial eigenvalues coincide within the stated numerical precision. For each fixed deformation direction \(d\), we evaluate the leading order-\(\rho^2\) splitting of the numerically unresolved fourth--fifth pair by treating its two-dimensional subspace with degenerate perturbation theory; the ordered fourth-eigenvalue shift is given by the larger perturbative eigenvalue (Appendix~\ref{app:E-spatial}). Time reflection and uniqueness of the conditional classical solution imply that the leading changes in these expansions are proportional to \(\rho^2\).

For \(d=\hat e_{\rm1L}\), numerical evaluation gives

\begin{equation}
R_2\simeq4.46>0,\qquad g_2\simeq0.986>0.
\label{eq:spatial-coefficients}
\end{equation} Both observables increase, indicating a modest strengthening of the existing ordered \(3+6\) spatial hierarchy. A mechanism for dynamical \(SO(9)\) to \(SO(3)\times SO(6)\) symmetry breaking has been proposed in a separate setting \cite{brandenberger2025}; the present calculation follows an existing hierarchy under prescribed local deformations.

Although the one-loop gradient distinguishes orientations within \(\mathcal T_{\rm soft}\), the quadratic conditional classical response is the same for every unit direction in this plane: \(R_2\) and \(g_2\) are independent of \(\varphi\).

The two matrices spanning this plane obey the algebraic relations of Pauli matrices on a two-dimensional subspace of background color space. Let \(e\) be the unit vector of the one-dimensional block from Section~\ref{sec:plane} and \(t\) the normalized complex-kernel vector in the five-dimensional color space, with \(e^\dagger t=0\). The trace-orthonormal matrices are

\begin{equation}
H_1:=\frac{et^\dagger+te^\dagger}{\sqrt2},\qquad
H_2:=\frac{-i et^\dagger+i te^\dagger}{\sqrt2}.
\label{eq:soft-plane-basis}
\end{equation} An arbitrary unit direction can be written as \(H(\varphi):=H_1\cos\varphi+H_2\sin\varphi\). With \(\Pi_{\rm c}:=ee^\dagger+tt^\dagger\),

\begin{equation}
H(\varphi)^2=\frac12\Pi_{\rm c}.
\end{equation} The square is therefore angle independent. Trace orthogonality to the spatial background also gives \(e^\dagger A_i^{(0)}t=0\), making \(H(\varphi)A_i^{(0)}H(\varphi)\) angle independent. Hence the double commutator sourcing the quadratic response in the 154 residual directions,

\begin{equation}
[H(\varphi),[H(\varphi),A_i^{(0)}]]
=\frac12\{\Pi_{\rm c},A_i^{(0)}\}
-2H(\varphi)A_i^{(0)}H(\varphi)
\end{equation} is angle independent. Fixed-scale normalization is also common, and the same invertible residual Hessian determines the quadratic spatial response throughout the plane. Therefore \(R_2\) and \(g_2\) are angle independent (Appendix~\ref{app:E-pauli}).

Thus the one-loop gradient distinguishes orientations at first order, whereas the leading conditional classical spatial response is identical for every unit direction in the plane. Angle independence follows analytically from the background algebra; the positive values of \(R_2\) and \(g_2\) are numerical properties of this background.

\begin{figure}[H]
\centering
\captionsetup[subfigure]{font=small,labelfont=bf,skip=9pt,justification=centering}
\begin{subfigure}{0.86\linewidth}
\centering
\includegraphics[width=\linewidth]{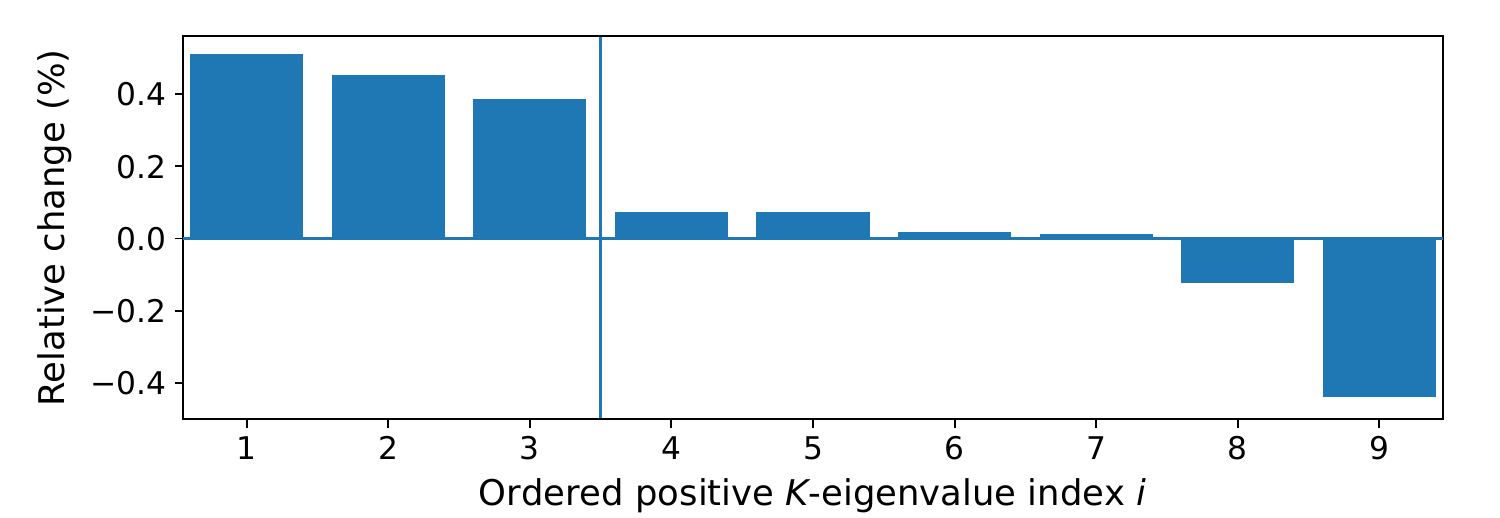}
\caption{Relative change of each positive \(K\)-eigenvalue at \(\rho=0.04\).}
\label{fig:spatial-panel-a}
\end{subfigure}
\par\vspace{8pt}
{\color{black!25}\rule{0.70\linewidth}{0.35pt}}
\par\vspace{7pt}
\begin{subfigure}{0.86\linewidth}
\centering
\includegraphics[width=\linewidth]{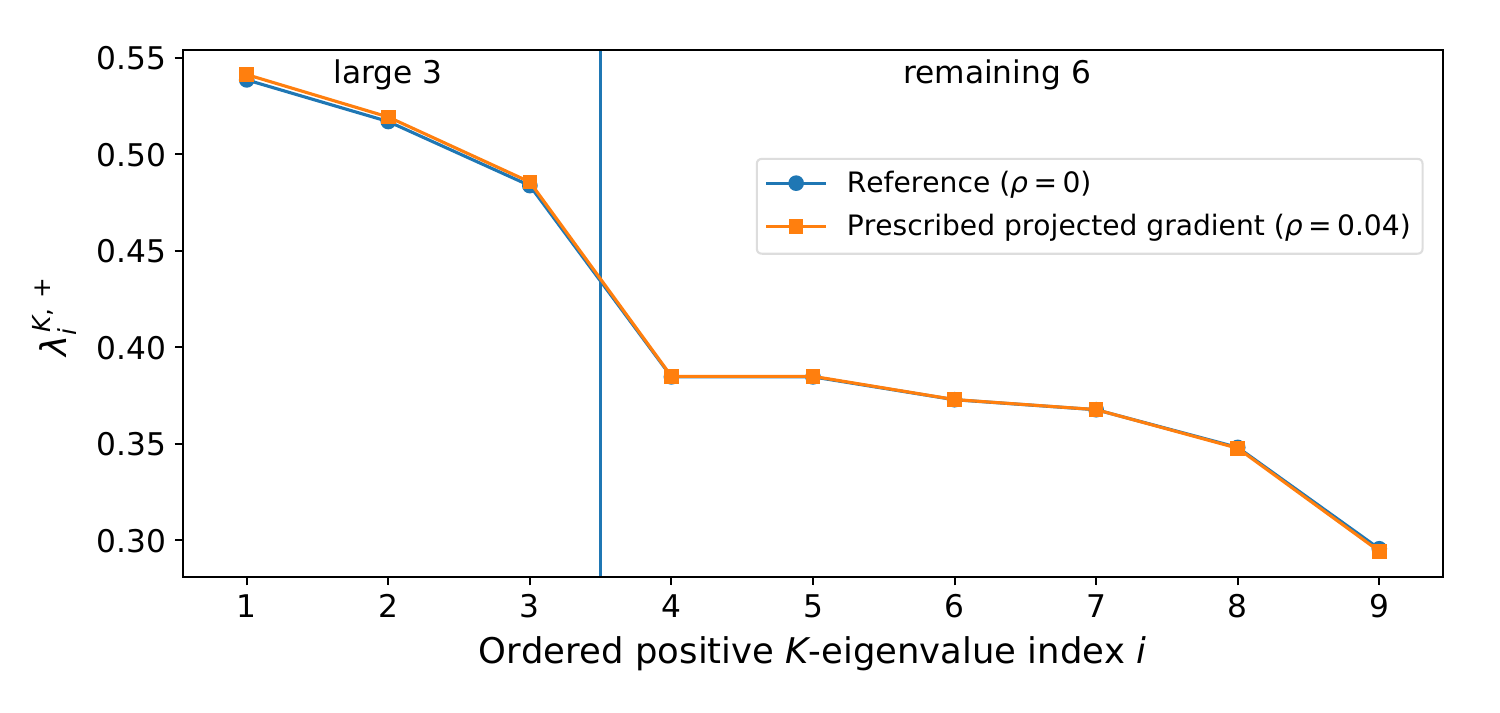}
\caption{Positive \(K\)-spectra at the reference point and at \(\rho=0.04\).}
\label{fig:spatial-panel-b}
\end{subfigure}
\par\vspace{8pt}
{\color{black!25}\rule{0.70\linewidth}{0.35pt}}
\par\vspace{7pt}
\begin{subfigure}{0.62\linewidth}
\centering
\includegraphics[width=\linewidth]{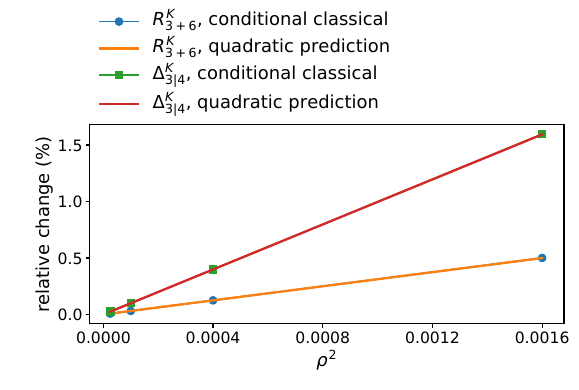}
\caption{Relative changes of the hierarchy ratio and the \(3|4\) gap.}
\label{fig:spatial-panel-c}
\end{subfigure}
\captionsetup{skip=13pt}
\caption{Conditional classical spatial response along \(\hat e_{\rm1L}\). (a) Subpercent relative changes of the individual eigenvalues at \(\rho=0.04\); (b) nearly overlapping reference and deformed spatial spectra; (c) modest changes in the hierarchy ratio and the \(3|4\) gap (markers: finite-amplitude numerical results; lines: quadratic expansion). Panels (a) and (b) have different vertical scales.}
\label{fig:spatial-response}
\end{figure}

Figure~\ref{fig:spatial-response}\subref{fig:spatial-panel-a} resolves the changes of the nine ordered positive \(K\)-eigenvalues at \(\rho=0.04\): the largest three increase, while the remaining six respond nonuniformly. Individual shifts range only from approximately \(-0.44\%\) to \(+0.51\%\), so the reference and deformed spectra nearly overlap in panel~\subref{fig:spatial-panel-b}. Panel~\subref{fig:spatial-panel-c} shows increases of approximately 0.5\% in the hierarchy ratio and 1.6\% in the third--fourth gap, consistent with the quadratic expansion. These small shifts quantify the conditional strengthening of the existing spatial hierarchy.

Uniform scale normalization cancels from the ratio coefficient \(R_2\) but contributes to the absolute-gap coefficient \(g_2\); temporal--spatial mixing \(G_{0i}=O(r^3)\) does not enter either coefficient at quadratic order. The remaining spatial response comes from classical stationarity in the 154 residual directions. The projector onto the leading three spatial directions also changes at quadratic order, showing mixing between the groups in the reference rest frame (Appendix~\ref{app:E-spatial}).

The positive response in this two-plane does not extend uniformly to all 15 temporal directions. Stationarity is imposed only in the 154 residual directions, so an action gradient can remain along the prescribed deformations. Full quantum stationary conditions and backreaction are future questions (Section~\ref{sec:discussion}).

\section{Conclusions and Discussion}\label{sec:discussion}

At finite \(N=5\), the internal geometry of a constrained Lorentzian IIB matrix background determines a distinguished real two-plane of temporal deformations. For equal-amplitude temporal deformations, the classical temporal extent is direction-independent through fourth order, whereas the logarithmic magnitude of the fermionic Pfaffian exhibits directional sensitivity already at first order. The lowest positive-eigenvalue doublet of the full fermionic operator is numerically most sensitive within this plane and dominates the first-order Pfaffian response. The identity linking the color--spin coupling vector, the temporal Hessian, and the doublet sensitivity connects the background geometry to the directional response of the fermionic weight.

Under the stated local reflection-covariance and nonsingularity conditions, the non-Pfaffian one-loop contributions cancel at first order when combined. Thus \(J_{\rm1L}=J_{\rm Pf}\) for the logarithmic magnitude of the retained-coordinate local density. Prescribing a temporal deformation along the in-plane projection of this gradient and imposing classical stationarity in the residual directions modestly strengthens the existing \(3+6\) spatial hierarchy. The quadratic spatial response is common to all unit directions in the plane, although the first-order Pfaffian response distinguishes their orientations. The two responses share a geometric origin while describing different operations.

The background construction is supported by an exact nearby constrained solution and an analytic local physical modulus. Its phase family preserves the quadratic geometry, the distinguished temporal plane and its scalar Hessian action, and the classical action. Rigorous pointwise bounds place the 23 numerical configurations of the selected branch close to this common analytic family after individual gauge and connected Lorentz transformations. This supports the classical background used in the local analysis; the full fermionic response is evaluated at the numerical reference configuration.

The next step is to study the full one-loop stationary conditions in both the temporal directions and the classical modulus, including quantum backreaction. A dynamical interpretation also requires the retained-coordinate integral with its Lorentzian phase and integration cycle, and comparison with other backgrounds and matrix sizes. Recent work on supersymmetry in Lorentzian simulations \cite{anagnostopoulosPoS2026} and on weak-coupling quantum matrix fluctuations \cite{steinacker2026} provides complementary settings for these questions.

\appendix
\numberwithin{equation}{section} 
\section{Physical quotient and analytical two-plane}\label{app:A}

We derive the conditional classical response from the splitting of the physical tangent space and determine the action of the Hessian on the two-dimensional real plane defined by the complex background kernel. The algebraic derivations use the rank, stationarity, and nondegeneracy conditions stated in each subsection.

\subsection{Constraint surface, orbit quotient, and physical Hessian}\label{app:A-quotient}

Let \(e_\ell\), \(\ell=1,\ldots,24\), be a trace-orthonormal Hermitian basis of \(\mathfrak{su}(5)\), and expand \(A_\mu=\sum_\ell a_\mu^\ell e_\ell\). We construct a local tangent space to the fixed-\(Q_\rho\) surface in the 240-dimensional real coefficient space by removing the color-gauge and Lorentz-frame directions. Equation~\eqref{eq:dimension-chain} uses rank 24 for the gauge tangent map, rank 1 for the constraint differential, and rank 45 for the Lorentz tangent map after quotienting.

Define the constrained Hessian at the reference point by

\begin{equation}
H_{\mathcal L}:=D^2(S_{B,0}-\lambda_Q Q_\rho)\big|_{A^{(0)}}.
\label{eq:A-lagrangian-hessian}
\end{equation}

The multiplier \(\lambda_Q\) is fixed to its value at the reference stationary point. The second variation of the action on the constraint surface includes the constraint force, so we use \(H_{\mathcal L}\), rather than restricting \(D^2S_{B,0}\) alone to the tangent space. Using the positive-definite coefficient inner product, we choose horizontal representatives and separate the modulus tangent. Denote the quadratic form on the 170-dimensional physical tangent space by \(\mathscr H\) and its 169-dimensional nonzero-curvature restriction by \(H_{169}\). Numerical analysis at the reference background gives \(\operatorname{Inertia}(\mathscr H)=(95,74,1)\), with the single null direction corresponding to the classical modulus. In our previous paper \cite{muramatsu2026}, the Lorentz tangent map after projection away from the gauge directions was numerically found to have rank 45 at all 23 sampled configurations along the selected classical solution family.

Numerical residuals and spectral quantities for the reference background are summarized below.

\begin{center}
\normalsize
\renewcommand{\arraystretch}{1.15}
\begin{tabular}{>{\raggedright\arraybackslash}p{0.46\linewidth}>{\raggedright\arraybackslash}p{0.43\linewidth}}
\toprule
Quantity & Numerical result\\
\midrule
$Q_\rho-9$ & Approximately $3.6\times10^{-15}$\\
Rest-frame $\|A_0\|_F$ & Approximately $4.4\times10^{-15}$\\
Relative residual of the nine spatial stationarity equations & Approximately $6.3\times10^{-16}$\\
15-dimensional temporal Hessian & 12 negative, 3 positive; smallest positive value $0.0431844482$\\
Spatial gap between the two groups, $\Delta_{3|4}^K$ & $0.0990229805$\\
Relative residuals of all fermionic eigenpairs & At most approximately $2.8\times10^{-15}$\\
\bottomrule
\end{tabular}
\end{center}
\medskip

Here \(\Delta X:=\sum_{i=1}^{9}[A_i^{(0)},[A_i^{(0)},X]]\). For the nine-matrix tuple \(X=(X_1,\ldots,X_9)\), we use the Frobenius norm \(\|X\|_F^2:=\sum_{i=1}^{9}\operatorname{Tr}(X_i^\dagger X_i)\); for a single matrix, the sum is omitted. The relative residual of the nine spatial stationarity equations is \(\|\Delta A-2\lambda_Q A\|_F/\|\Delta A\|_F\), where \(A=(A_1^{(0)},\ldots,A_9^{(0)})\) and \(\Delta A=(\Delta A_1^{(0)},\ldots,\Delta A_9^{(0)})\). These numerical results concern the reference background. For the nearby exact solution, Appendix~\ref{app:B-certificate} establishes a rigorous bound on the 84-dimensional transverse, nondegenerate Hessian within the static reflection sector, with inertia \((49,35,0)\). This sector is distinct from the full 170-dimensional physical Hessian.

\subsection{Hessian-adapted splitting and conditional classical response}\label{app:A-splitting}

Let the kernel of the Hessian on the 170-dimensional physical tangent space be \(\mathbb Rt_u\). If the quadratic form on the temporal-deformation space \(W\) is nondegenerate, then \(W\cap\ker\mathscr H=\{0\}\), so the 15 temporal deformations embed in the 169-dimensional space with the modulus direction removed. Let \(Q_{\rm time}\) be the embedding obtained from representatives orthogonal to \(t_u\) in the positive-definite coefficient inner product. The Hessian-orthogonal complement

\begin{equation}
\mathcal V_{\rm res}:=\ker(Q_{\rm time}^{\mathsf T}H_{169}),
\qquad \dim\mathcal V_{\rm res}=154
\label{eq:A-residual-space}
\end{equation}
then forms a direct sum with the temporal-deformation space. Writing its basis columns as \(Z_{\rm res}\), an invertible change of basis gives

\begin{equation}
[Q_{\rm time},Z_{\rm res}]^{\mathsf T}H_{169}
[Q_{\rm time},Z_{\rm res}]
=\begin{pmatrix}H_{\rm time}&0\\0&H_{\rm res}\end{pmatrix}.
\label{eq:A-block-split}
\end{equation}

This orthogonality refers to the Hessian quadratic form. Sylvester's law of inertia gives \((95,74,0)=(3,12,0)+(92,62,0)\), so the residual block is indefinite and invertible.

Let \(E_{\mu p}\), \(p=1,\ldots,154\), be the residual tangent vectors in the rest frame, and choose horizontal representatives with \(E_{0p}=0\). Appendix~\ref{app:C-reflection} explains why background-preserving time reflection and nondegeneracy of the temporal block permit this choice. Define the unnormalized configuration and its map to fixed scale by

\begin{equation}
\begin{gathered}
B_0(c,y)=\sum_{a=1}^{15}c^aH_a,\qquad
B_i(c,y)=A_i^{(0)}+\sum_{p=1}^{154}y^pE_{ip},\\
q_{\rm raw}(c,y)=\operatorname{Tr}(B_\mu B^\mu),\qquad
A_\mu(c,y)=\sqrt{q_0/q_{\rm raw}(c,y)}\,B_\mu(c,y).
\end{gathered}
\label{eq:A-local-chart}
\end{equation}

Since the residual columns are tangent to the fixed-scale surface, they obey \(\sum_i\operatorname{Tr}(A_i^{(0)}E_{ip})=0\). The cross term between the background and the residual response therefore vanishes. In a neighborhood with \(q_{\rm raw}>0\), this map is smooth and preserves \(Q_\rho=9\). Set \(\mathcal L_{\rm loc}(c,y):=S_{B,0}[A(c,y)]\) and write the residual stationarity equations as \(f(c,y):=\partial_y\mathcal L_{\rm loc}(c,y)=0\). At an exact constrained stationary base, this local chart lies on the fixed-\(Q_\rho\) surface, so the Hessian of \(\mathcal L_{\rm loc}\) at the origin is the pullback of the constrained Hessian \(H_{\mathcal L}\) defined in Eq.~\eqref{eq:A-lagrangian-hessian}. Thus \(f(0,0)=0\), and the first-order expansion is

\begin{equation}
f(c,y)=H_{\rm mix}c+H_{\rm res}y
+O\bigl((\|c\|+\|y\|)^2\bigr),
\qquad H_{\rm mix}=0.
\label{eq:A-residual-expansion}
\end{equation}

On the section \(u=0\), these assumptions and the invertibility of \(H_{\rm res}\) give a unique smooth local solution of the 154 residual stationarity equations, with \(D_cy(0)=-H_{\rm res}^{-1}H_{\rm mix}=0\). This yields Eq.~\eqref{eq:quadratic-response}. At the numerical reference background, the 169-dimensional nonzero-curvature Hessian and the 154-dimensional residual block are nondegenerate according to their computed inertias. The computer-assisted proof in Appendix~\ref{app:B} establishes a nearby exact classical solution and a rigorous nondegeneracy bound for the 84-dimensional transverse Hessian within the static reflection sector.

\subsection{Invariance of the complex kernel and scalar Hessian action}\label{app:A-plane}

In a frame adapted to the background color reflection of the main text, write the three odd and six even spatial matrices as

\begin{equation}
A_a=\begin{pmatrix}0&z_a^\dagger\\z_a&0\end{pmatrix}
\quad(a=1,2,3),\qquad
A_b=\begin{pmatrix}a_b&0\\0&B_b\end{pmatrix}
\quad(b=4,\ldots,9).
\label{eq:A-color-blocks}
\end{equation}

Here \(B_b\) is Hermitian, and we assume that \(\mathsf Z=[z_1,z_2,z_3]\) has complex rank 3. The trace inner product of odd Hermitian temporal deformations \(c(w)\) is \(\operatorname{Tr}[c(w)c(v)]=2\operatorname{Re}(w^\dagger v)\). Orthogonality to the background is \(\operatorname{Re}\mathsf Z^\dagger w=0\), leaving five real dimensions of odd physical temporal deformations. Within this space, the complex-orthogonality condition \(\mathsf Z^\dagger w=0\) defines the two-dimensional real plane of Eq.~\eqref{eq:kernel-plane}.

Split the color Laplacian as \(\Delta=\sum_{i=1}^{9}[A_i,[A_i,\,\cdot\,]]=\Delta_3+\Delta_6\). With \(\tau_3=\operatorname{Tr}(\mathsf Z^\dagger\mathsf Z)\), direct block multiplication gives

\begin{equation}
\Delta_3c(w)=c\!\left(\tau_3w+\mathsf Z\mathsf Z^\dagger w
-2\mathsf Z\overline{\mathsf Z^\dagger w}\right).
\label{eq:A-three-laplacian}
\end{equation}

Let \(c(w_\alpha)\) be a trace-orthonormal basis of the odd physical temporal deformations, and set \(m_\alpha=\mathsf Z^\dagger w_\alpha\). Since \(\operatorname{Re}m_\alpha=0\),

\begin{equation}
(L_3)_{\alpha\beta}
:=\operatorname{Tr}[c(w_\alpha)\Delta_3c(w_\beta)]
=\tau_3\delta_{\alpha\beta}
+6\operatorname{Re}(m_\alpha^\dagger m_\beta).
\label{eq:A-three-gram}
\end{equation}

Thus, on the five-dimensional real reflection-odd physical deformation space, \(L_3-\tau_3I\) is positive semidefinite. Its kernel is precisely the two-dimensional real plane \(P_0\), defined by \(\mathsf Z^\dagger w=0\).

The combined action of the six matrices is \(\Delta_6c(w)=c(K_6w)\), with \(K_6=\sum_{b=4}^{9}(B_b-a_bI_4)^2\). Applying the fixed-\(Q_\rho\) stationarity equations of the background in its rest frame to the three odd components and setting \(G_Z:=\mathsf Z^\dagger\mathsf Z\), we obtain

\begin{equation}
K_6\mathsf Z
=\mathsf Z\bigl[(2\lambda_Q-\tau_3)I_3-G_Z+2G_Z^{\mathsf T}\bigr].
\label{eq:A-six-invariance}
\end{equation}

Hence, \(\operatorname{Ran}\mathsf Z\) and its orthogonal complement \(\ker\mathsf Z^\dagger\) are invariant subspaces of the Hermitian combined operator \(K_6\). On the one-complex-dimensional kernel, \(K_6z_0=\kappa_6z_0\), with \(\kappa_6\in\mathbb R\).

Let \(\mathcal E:\mathbb R^{15}\to\mathbb R^{24}\) denote the isometric embedding of the temporal deformation space into the real color-coefficient space, defined using the trace-orthonormal basis. The constrained temporal Hessian is then

\begin{equation}
H_{\rm time}=2\lambda_QI_{15}-\mathcal E^{\mathsf T}\Delta\mathcal E.
\label{eq:A-time-hessian}
\end{equation}

The scalar actions of \(\Delta_3\) and \(\Delta_6\) on \(P_0\) imply Eq.~\eqref{eq:plane-curvature} in the main text. The existence of the two-plane and the scalar action follow algebraically, whereas the small positive value of \(h_0\) and its identification as the lowest positive-curvature doublet at the selected background are numerical results in the specified normalization.

\section{Exact nearby classical solution and analytic phase modulus}\label{app:B}

This appendix gives the constrained classical construction underlying the background comparison in Section~\ref{sec:background} and the geometry in Section~\ref{sec:plane}. We present the equations and rigorous bounds of the computer-assisted existence proof and the pointwise comparison of 23 numerical configurations, then derive the phase family and its invariants. The contraction argument below determines the exact base point within the stated local enclosure.

\subsection{Static reflection sector and exact nearby existence}\label{app:B-existence}

Work in a fixed Lorentz rest frame with \(A_0=0\), \(R_0=\operatorname{diag}(-1,I_4)\), and the blocks of Eq.~\eqref{eq:A-color-blocks}. Write the odd matrices as \(X(z_a)\) and the even ones as \(Y_b=\operatorname{diag}(a_b,B_b)\), where \(a_b+\operatorname{Tr}B_b=0\). The color-odd and color-even spaces have real dimensions 8 and 16, respectively. The static reflection-fixed space therefore has \(3\times8+6\times16=120\) real coordinates \(x\), chosen in an exactly trace-orthonormal basis. Its residual continuous symmetry group is
\begin{equation}
\mathcal G_{\rm ref}=S(U(1)\times U(4))\times SO(3)\times SO(6),
\qquad \dim\mathcal G_{\rm ref}=16+3+15=34.
\label{eq:B-residual-group}
\end{equation}
Here \(R_0\) and \(-R_0\in SU(5)\) have the same conjugation action. Allowing a reflection-even temporal matrix would instead give 136 coordinates; that larger space is not the sector used in this proof.

Let \(E(x)\) be the coordinate vector of \(E_i=\sum_{j=1}^{9}[A_j,[A_j,A_i]]\). With fixed slice columns \(T\in\mathbb R^{120\times34}\), a fixed phase-slice vector \(v\in\mathbb R^{120}\), and slice origin \(x_c\), introduce \(\xi=(x,\lambda_Q,\beta,\alpha)\), with \(\beta\in\mathbb R^{34}\) and \(\alpha\in\mathbb R\). The square system is
\begin{equation}
F_{u_{\rm cert}}(\xi):=
\begin{pmatrix}
E(x)-2\lambda_Qx+T\beta+\alpha v\\
x^{\mathsf T}x-q_0\\
T^{\mathsf T}(x-x_c)\\
v^{\mathsf T}(x-x_c)-u_{\rm cert}
\end{pmatrix}=0,
\qquad q_0=15+2\sqrt3.
\label{eq:B-square-system}
\end{equation}
There are \(120+1+34+1=156\) equations and unknowns. The slice data \(T,v,x_c\), the rational center \(\widehat\xi\), and the inverse candidates are specified exactly in the separate verification data; the algebraic constants in the trace basis and \(q_0\) are enclosed rather than replaced by rounded constants. The coordinate \(u_{\rm cert}\) labels this proof's slice and is distinct from the retained numerical coordinate \(u\).

The added multipliers vanish exactly, not merely within a small numerical residual. To see the additional identity needed for this conclusion, set
\begin{equation}
\begin{gathered}
H_Z:=\mathsf Z^\dagger\mathsf Z=G_Z^{\rm R}+iC_Z,
\quad (G_Z^{\rm R})^{\mathsf T}=G_Z^{\rm R},\quad C_Z^{\mathsf T}=-C_Z,
\quad \tau_3=\operatorname{Tr}H_Z,\\
b:=((C_Z)_{23},-(C_Z)_{13},(C_Z)_{12})^{\mathsf T},
\qquad P_b:=bb^{\mathsf T}.
\end{gathered}
\label{eq:B-axis-data}
\end{equation}
Direct block multiplication gives the odd part of the equations,
\begin{equation}
E_Z=K_6\mathsf Z+\mathsf Z(H_Z+\tau_3I_3-2H_Z^{\mathsf T}).
\label{eq:B-odd-equation}
\end{equation}
The transpose in this expression is essential. Let \(w(x)\) be the real-coordinate vector for \(\delta\mathsf Z=i\mathsf ZP_b\), \(\delta Y_b=0\). Since \(C_Zb=0\), one has \([C_Z,P_b]=0\), and
\begin{equation}
\begin{aligned}
x^{\mathsf T}w&=2\operatorname{Re}\!\left[i\operatorname{Tr}(P_bH_Z)\right]=0,\\
w^{\mathsf T}E&=4\operatorname{Tr}\!\left(P_b[G_Z^{\rm R},C_Z]\right)
=4\operatorname{Tr}\!\left([C_Z,P_b]G_Z^{\rm R}\right)=0.
\end{aligned}
\label{eq:B-offshell-identity}
\end{equation}
These are off-shell identities and use neither stationarity nor division by \(b^{\mathsf T}b\). Together with the 34 Noether identities for the orbit columns \(V_{\rm ref}(x)\), they give, at a zero of Eq.~\eqref{eq:B-square-system},
\begin{equation}
\mathcal B(x)\binom{\beta}{\alpha}=0,
\qquad
\mathcal B(x):=[V_{\rm ref}(x),w(x)]^{\mathsf T}[T,v].
\label{eq:B-border-closure}
\end{equation}
The computer-assisted verification gives the uniform bound \(\|I-R_{\mathcal B}\mathcal B(x)\|_\infty<2.068118\times10^{-6}<1\) throughout the wide box defined below, for a fixed inverse candidate \(R_{\mathcal B}\). Thus \(\mathcal B\) is invertible and \(\beta=\alpha=0\). Equivariance under the combined color and spatial reflection puts the spatial gradient in the 120-dimensional fixed space. The omitted spatial components therefore vanish as well, and the temporal equation vanishes because \(A_0=0\). The resulting solution satisfies Eq.~\eqref{eq:constrained-EOM} in all ten matrix components.

\subsection{Enclosure, rank bounds, and the local physical branch}\label{app:B-certificate}

For an exact rational preconditioner \(R_N\), put
\begin{equation}
\begin{gathered}
\eta_N:=\|R_NF_0(\widehat\xi)\|_\infty,\qquad
q_N:=\|I-R_NDF(\widehat\xi)\|_\infty,\\
q_N(r):=q_N+16000\|R_N\|_\infty r.
\end{gathered}
\label{eq:B-newton-bounds}
\end{equation}
On \(\|\xi-\widehat\xi\|_\infty\le r\le2^{-10}\), with \(\|\widehat x\|_2<5\), the commutator inequality \(\|[A,B]\|_F\le2\|A\|_F\|B\|_F\) yields
\begin{equation}
\|DF(\xi)-DF(\widehat\xi)\|_\infty\le16000r.
\label{eq:B-lipschitz}
\end{equation}
For example, the cubic-equation rows are bounded by \((14524+15972r)r\), and the quadratic constraint row by \(240r\); the slice rows are constant. The map \(\xi\mapsto\xi-R_NF_{u_{\rm cert}}(\xi)\) is a strict contraction into the box if
\begin{equation}
q_N(r)<1,\qquad
\eta_N+\|R_Ne_{156}\|_\infty|u_{\rm cert}|+q_N(r)r<r.
\label{eq:B-inclusion}
\end{equation}
Since \(q_N<1\) also makes \(R_N\) invertible, its fixed point solves the original square system. Independent 448-bit outward-rounded integer-endpoint verification gives the following conservative bounds:
\begin{center}
\begin{tabular}{@{}ll@{}}
\toprule
Quantity & Certified bound or domain\\
\midrule
\(\eta_N\) & \(<2.784\times10^{-99}\)\\
\(q_N\) & \(<5.619\times10^{-13}\)\\
\(\|R_N\|_\infty\) & \(<35.201\)\\
Narrow box, \(u_{\rm cert}=0\) & \(r_s=2^{-250}\); image radius \(<3.106\times10^{-88}<r_s\)\\
Wide box and slice domain & \(r_w=2^{-28}\), \(|u_{\rm cert}|\le2^{-32}\)\\
Wide-box contraction and image & \(q_N(r_w)<0.002098128\); image radius \(<1.390\times10^{-10}<r_w\)\\
\bottomrule
\end{tabular}
\end{center}
The displayed decimal bounds summarize the integer-endpoint inequalities; the exact centers, inverse candidates, and endpoints are separate verification data, not reconstructed from these rounded displays.

The same box has \(2.77<\lambda_Q<2.78\), \(b^{\mathsf T}b>0\), and \(\operatorname{rank}_{\mathbb C}\mathsf Z=3\). Verified positive-definite factorizations give
\begin{equation}
\bigl(\operatorname{Tr}A_iA_j\bigr)_{i,j=1}^{9}>1.47I_9,
\qquad \Delta>5.4I_{24}.
\label{eq:B-rank-bounds}
\end{equation}
The first Gram matrix is unnormalized: the moment matrix of Eq.~\eqref{eq:moments-G} contains the additional factor \(1/5\). The second bound is on traceless color matrices and excludes a nonzero common commutant, so the background is irreducible and its color orbit has rank 24. A separately verified inverse bound \(\|I-R_sV_{\rm ref}^{\mathsf T}T\|_\infty<1.738\times10^{-6}\) establishes orbit rank 34 and transverse slicing. A verified congruence of the bordered Hessian has inertia \((85,71,0)\). Subtracting the 36 positive--negative pairs associated with the constraint and slice borders gives
\begin{equation}
120-34-1-1=84,\qquad
\operatorname{Inertia}(H_{\rm hard}^{\rm ref})=(49,35,0),\qquad
\min|\lambda(H_{\rm hard}^{\rm ref})|>0.04599.
\label{eq:B-hard-inertia}
\end{equation}
These 84 nonzero-curvature directions (``hard'' in \(H_{\rm hard}^{\rm ref}\)) are transverse to symmetry orbits and the modulus in the static reflection sector; they differ from the 170-dimensional physical fluctuation space and the 154-dimensional one-loop residual block.

The analytic implicit-function theorem now gives a real-analytic slice solution for \(|u_{\rm cert}|\le2^{-32}\), locally about each parameter value. It has \(T^{\mathsf T}x'=0\) and \(v^{\mathsf T}x'=1\). Transversality of \(T^{\mathsf T}V_{\rm ref}\) excludes \(x'\) being an orbit tangent. Decomposing the full color and spatial rotation Lie algebras by the reflection shows that their tangents in this fixed space are generated by \(\mathcal G_{\rm ref}\). A nonzero boost has a temporal component at a rank-9 static background. Hence this slice defines a nontrivial local physical modulus, rather than a change of color or Lorentz frame.

For the distance estimate, let \(\widehat A_{\rm cert}\) denote the numerical reference used in the existence proof and \(\widehat A_{23}\) the reference configuration (index 0) of the earlier 23-point numerical continuation. The numerical color and spatial frame matrices are replaced by their exact polar factors, and the boost is constructed from its recorded parameters before transforming the enclosed root back to the original frame. Including these corrections gives
\begin{equation}
\|A_*-\widehat A_{\rm cert}\|_{F,+}<5.067172\times10^{-14}<5.1\times10^{-14}.
\label{eq:B-native-distance}
\end{equation}
Here \(\|A\|_{F,+}^2:=\sum_{\mu=0}^{9}\operatorname{Tr}(A_\mu^\dagger A_\mu)\) is an auxiliary positive-definite tuple norm. The separate comparison of the two reference arrays gives \(\|\widehat A_{23}-\widehat A_{\rm cert}\|_{F,+}<1.43\times10^{-15}\). Combining this with Eq.~\eqref{eq:B-native-distance} yields the conservative bounds
\begin{equation}
\|A_*-\widehat A_{\rm cert}\|_{F,+}<5.1\times10^{-14},
\qquad
\|A_*-\widehat A_{23}\|_{F,+}<5.22\times10^{-14}.
\label{eq:nearby-root-bounds}
\end{equation}
These two numerical reference configurations are not identical. The 15-direction Pfaffian response is evaluated at the same numerical reference background \(\widehat A_{\rm cert}\), for which the existence of a nearby exact classical solution has been established. The classical proximity bound alone does not provide error bounds for the fermionic spectrum or Pfaffian derivatives at the exact solution.

A stronger pointwise comparison uses the common exact phase family of Appendix~\ref{app:B-phase}. Write this family in a fixed reflection-adapted frame as \(A_*^{\rm ph}(\theta_{\rm mod})\), and denote the full ten-matrix numerical configurations in continuation order by \(\widehat A^{(i)}\), \(i=0,\ldots,22\). For each configuration, the rigorous pointwise bound provides a phase \(\theta_{{\rm mod},i}\), color transformation \(g_i\in SU(5)\), and proper, orthochronous Lorentz transformation \(\Lambda_i\in SO^+(9,1)\), such that
\begin{equation}
\left\|\widehat A^{(i)}-g_i\cdot\bigl(\Lambda_i\cdot A_*^{\rm ph}(\theta_{{\rm mod},i})\bigr)\right\|_{F,+}
<1.65\times10^{-14},\qquad i=0,\ldots,22.
\label{eq:B-pointwise-bridge}
\end{equation}
The norm includes all ten matrices without normalization. These 23 configurations sample one continued family, rather than the 23 distinct classes found in the broader search. The transformations and phases are determined and verified individually for each configuration, so the comparison uses a different alignment from the direct reference-configuration bounds above. The resulting bounds establish pointwise proximity to a common analytic family; they do not establish exact stationarity of the numerical configurations, uniqueness of a global continuum branch, or error bounds that transfer the numerical Pfaffian and eigenspace results to the exact solutions.

\subsection{Explicit phase curve and preservation of the equations}\label{app:B-phase}

At a stationary solution, \(\mathsf Z^\dagger E_Z=2\lambda_QH_Z\) is Hermitian. Equation~\eqref{eq:B-odd-equation} therefore implies
\begin{equation}
[H_Z,H_Z^{\mathsf T}]=0
\quad\Longleftrightarrow\quad [G_Z^{\rm R},C_Z]=0.
\label{eq:B-commuting-gram}
\end{equation}
As \(b^{\mathsf T}b>0\), define the real rank-one projector
\begin{equation}
P_{\rm ax}:=\frac{bb^{\mathsf T}}{b^{\mathsf T}b},\qquad
U_{\rm mod}:=e^{i\theta_{\rm mod}P_{\rm ax}}.
\label{eq:B-axis-projector}
\end{equation}
A real symmetric matrix commuting with the nonzero antisymmetric \(C_Z\) preserves its axis and is scalar on the orthogonal two-plane. Thus \(P_{\rm ax}\) commutes with both \(H_Z\) and \(H_Z^{\mathsf T}\). Starting from the exact solution \(A_*\), define the phase curve by
\begin{equation}
\mathsf Z(\theta_{\rm mod})=\mathsf Z_*e^{i\theta_{\rm mod}P_{\rm ax}},
\qquad Y_b(\theta_{\rm mod})=Y_{b,*},\qquad A_0(\theta_{\rm mod})=0.
\label{eq:analytic-phase-branch}
\end{equation}
Along this curve,
\begin{equation}
\mathsf Z^\dagger\mathsf Z=H_{Z,*},\qquad
\mathsf Z\mathsf Z^\dagger=\mathsf Z_*\mathsf Z_*^\dagger.
\label{eq:B-two-grams}
\end{equation}
The first identity uses commutation as well as unitarity; the second uses unitarity alone. Equation~\eqref{eq:B-odd-equation} consequently transforms by right multiplication by \(U_{\rm mod}\). For each even matrix, the sum of odd contributions to the lower block of its equation is
\begin{equation}
\sum_{a=1}^{3}[X(z_a),[X(z_a),Y_b]]_{22}
=\{\mathsf Z\mathsf Z^\dagger,B_b\}-2a_b\mathsf Z\mathsf Z^\dagger,
\label{eq:B-even-equation}
\end{equation}
and the upper block is fixed by tracelessness. Hence the even equations also remain unchanged. Fixed \(Y_b\) and Eq.~\eqref{eq:B-two-grams} preserve \(Q_\rho\), so the constrained equations hold with the same \(\lambda_Q\) for every real \(\theta_{\rm mod}\).

Locally returning this curve to the transverse slice gives a nonzero derivative. With the notation above, the derivative satisfies
\begin{equation}
\frac{du_{\rm cert}}{d\theta_{\rm mod}}
=\frac{w^{\mathsf T}v-w^{\mathsf T}T(V_{\rm ref}^{\mathsf T}T)^{-1}V_{\rm ref}^{\mathsf T}v}{b^{\mathsf T}b},
\qquad
0.9437236071<\frac{du_{\rm cert}}{d\theta_{\rm mod}}<0.9437240264
\label{eq:B-phase-slice-speed}
\end{equation}
at the base point \(\theta_{\rm mod}=0\), corresponding to \(u_{\rm cert}=0\). Thus the phase parameter traces the same local branch of physically inequivalent classical solutions obtained from the square system. This derivative is not a conversion from the old coordinate \(u\); the interval \(|u_{\rm cert}|\le2^{-32}\) is neither an interval in \(u\) nor a certified \(\theta_{\rm mod}\) coverage interval. The pointwise bounds in Eq.~\eqref{eq:B-pointwise-bridge} place all 23 numerical configurations close to this common family after the respective transformations. These bounds do not identify the whole numerical continuation with a globally unique analytic branch.

\subsection{Geometric and action invariants}\label{app:B-invariants}

In the same fixed reflection-adapted frame, the remaining quadratic traces are
\begin{equation}
\begin{gathered}
\operatorname{Tr}X(z_a)X(z_c)=2\operatorname{Re}(H_Z)_{ac},\qquad
\operatorname{Tr}X(z_a)Y_b=0,\\
\operatorname{Tr}Y_bY_d=a_ba_d+\operatorname{Tr}(B_bB_d),\qquad G_{0\mu}=0.
\end{gathered}
\label{eq:B-full-gram}
\end{equation}
Together with Eq.~\eqref{eq:B-two-grams}, these identities preserve every component of \(G_{\mu\nu}\), and hence the spectrum of \(K\), the hierarchy ratio, and the third--fourth gap wherever those observables are defined. A parameter-dependent change of Lorentz frame may change the components of the Gram matrix while leaving the spectrum of \(K\) invariant; a positive gap bound requires a separate enclosure.

Invertibility of \(U_{\rm mod}^\dagger\) gives \(\ker\mathsf Z(\theta_{\rm mod})^\dagger=\ker\mathsf Z_*^\dagger\). Complex rank 3 makes this kernel a complex line and defines the same ambient real two-plane \(P_0\) of temporal matrix deformations at every phase. Although the full 15-dimensional temporal complement can move, this plane lies in it throughout the curve. Stationarity in Eq.~\eqref{eq:A-six-invariance} and Hermiticity of \(K_6\) make the kernel invariant under \(K_6\). Since the kernel and \(K_6\) are fixed, its scalar eigenvalue \(\kappa_6\) is fixed. The already established scalar action then gives
\begin{equation}
h_0(\theta_{\rm mod})=2\lambda_Q-\tau_3-\kappa_6=h_0(0).
\label{eq:B-fixed-curvature}
\end{equation}
The positivity and lowest-pair ordering of this scalar curvature are separate numerical properties of the reference background. A unit generator of the fixed kernel also keeps \(F_z,H_z,W_z\) and \(\mathcal Z_F\) fixed, while its coupling to the complementary full fermionic space can vary.

Finally, along a fixed-scale stationary family,
\begin{equation}
\frac{dS_{B,0}}{d\theta_{\rm mod}}
=\lambda_Q\frac{dQ_\rho}{d\theta_{\rm mod}}=0,
\qquad 4S_{B,0}=2\lambda_Qq_0.
\label{eq:B-action-invariant}
\end{equation}
Thus the classical action, also including its fixed factor \(g^{-2}\), is constant. Differentiating the fixed-\(\lambda_Q\) equations gives a constrained Hessian zero vector with \(\dot{\mathsf Z}=i\mathsf Z_*P_{\rm ax}\), \(\dot Y_b=0\), and \(\dot A_0=0\). The local transversality argument makes its physical-quotient class nonzero. This establishes a zero direction, not unique nullity of the full physical Hessian. Exact reflection permits the complete-projector selection rule of Appendix~\ref{app:D-sensitivity} on this branch, but none of these invariants fixes the full fermion spectrum, the Pfaffian, its two-plane fractions, the local one-loop density, or the conditional-response coefficients along \(\theta_{\rm mod}\).

\section{Lorentzian one-loop measure and reflection cancellation}\label{app:C}

\subsection{Fresnel boundary value and fermionic conventions}\label{app:C-fresnel}

Keeping the 15 temporal-deformation coefficients \(c\) and the modulus coordinate fixed, we integrate over quadratic fluctuations in the 154 residual directions around the conditional classical configuration \(A(c,y(c))\). For a real symmetric nondegenerate quadratic form \(H\) and a positive scale \(\tau\), define the Lorentzian Gaussian as the Abel-damped boundary value

\begin{equation}
\begin{aligned}
\int_{\rm Fr}d^nx\,e^{ix^{\mathsf T}Hx/(2\tau)}
&:=\lim_{\epsilon\downarrow0}\int_{\mathbb R^n}d^nx\,
 e^{ix^{\mathsf T}Hx/(2\tau)-\epsilon\|x\|_+^2}\\
&=(2\pi\tau)^{n/2}
\frac{e^{i\pi\operatorname{sig}H/4}}{\sqrt{|\det H|}},
\qquad \operatorname{sig}H=n_+(H)-n_-(H).
\end{aligned}
\label{eq:C-fresnel}
\end{equation}

Here \(\|x\|_+^2\) is an auxiliary positive-definite norm on the finite-dimensional fluctuation coordinates, distinct from the spacetime Lorentzian metric. This damping defines the oscillatory boundary value while retaining the inertia and Fresnel phase of the Lorentzian quadratic form. The same prescription is used for the local integral over the 154 residual directions.

The quadratic form \(H\) used here is the constrained second variation of \(S_{B,0}\) pulled back to the same local coordinates. Thus, in the convention \(\exp(iS_{B,0}/g^2)\) with \(\hbar=1\), we have \(\tau=g^2\); restoring \(\hbar\) gives \(\tau=\hbar g^2\). This common factor is independent of \(c\) and does not contribute to the first-order gradient of the logarithmic magnitude at fixed \(g\).

In a trace-orthonormal color basis, \(D_X=i[X,\,\cdot\,]\) is real antisymmetric. We use real symmetric same-chirality bilinear blocks \(\widehat\gamma^\mu\), with \(\widehat\gamma^0=I_{16}\) and \(\{\widehat\gamma^i,\widehat\gamma^j\}=2\delta^{ij}I_{16}\) for the spatial components. Thus, \(\mathcal M_F\) in the main text is real antisymmetric, \(\mathcal K_F=i\mathcal M_F\) is Hermitian, and its nonzero eigenvalues form positive--negative pairs. The ordered Majorana--Weyl and color bases and the Berezin orientation are held fixed throughout all comparisons. In a local region where the Pfaffian is nonzero, \(\operatorname{Pf}(\mathcal M_F)^2=\det\mathcal M_F\) implies the inverse-matrix trace representation of the logarithmic derivative in Eq.~\eqref{eq:local-response}.

\subsection{Local measure and determinant of the full coordinate frame}\label{app:C-measure}

In the 240-dimensional coefficient space, set \(\mathsf J:=\eta\otimes I_{24}\), \(q_{\rm fix}=A^{\mathsf T}\mathsf JA=q_0\), and \(\nu:=2\mathsf JA\). With respect to the auxiliary positive-definite coefficient inner product, \(\nu\) is the normal to the constraint surface. The differential of the map from unnormalized coefficients \(b\) to the fixed-scale surface is

\begin{equation}
A=\alpha b,\quad
\alpha=\left(\frac{q_{\rm fix}}{b^{\mathsf T}\mathsf Jb}\right)^{1/2},
\qquad
j_b(F)=\alpha\left(F-b\frac{b^{\mathsf T}\mathsf JF}{b^{\mathsf T}\mathsf Jb}\right).
\label{eq:C-scale-differential}
\end{equation}

Let \(\mathsf O\) contain the 15 temporal-deformation columns, \(\mathsf E_h\) the 154 Hessian-orthogonal residual columns, and \(v_u\) be a horizontal representative of the modulus tangent. Set \(h:=j_b(\mathsf E_h)\) and \(\mathsf V_{\rm ret}:=j_b([v_u,\mathsf O])\). The retained frame \(\mathsf V_{\rm ret}\) combines one modulus column and 15 temporal-deformation columns.

Let \(\mathsf V_g\) be the 24-column color-gauge orbit frame, \(L\) the full 45-column Lorentz-orbit frame, and \(U=[\mathsf V_g,L]\). In the auxiliary positive-definite coefficient inner product, let \(P_g\) project out \(\mathsf V_g\) and \(P\) project out \([U,\nu]\). Define

\begin{equation}
\begin{gathered}
G_g=\mathsf V_g^{\mathsf T}\mathsf V_g,\qquad G_f=L^{\mathsf T}P_gL,
\qquad X=Ph,\qquad D=X^{\mathsf T}X,\\
G_r=\mathsf V_{\rm ret}^{\mathsf T}P\mathsf V_{\rm ret}-\mathsf V_{\rm ret}^{\mathsf T}XD^{-1}X^{\mathsf T}\mathsf V_{\rm ret},
\qquad H_{154}=D^{-1/2}H_cD^{-1/2}.
\end{gathered}
\label{eq:C-gram-blocks}
\end{equation}

Here \(G_f\) is the Lorentz Gram matrix after the color-gauge quotient, and \(G_r\) is the Schur--Gram matrix of the retained frame with mixing with the residual columns removed. The matrix \(H_c\) is the constrained Hessian in the residual coordinate basis before orthonormalization, and \(H_{154}\) represents the same quadratic form in an orthonormal residual frame. We assume below that all the Gram matrices, \(H_c\), and the full coordinate frame are nonsingular.

Let \(\Delta_{\rm FP}\) denote the Faddeev--Popov matrix for the same background gauge condition. In the orbit-adapted local coordinates used here, the quotient measure contains the inverse of the explicit Faddeev--Popov determinant, as follows from the gauge-slice identity in our previous paper \cite{muramatsu2026}. Up to deformation-independent constants, the local measure is

\begin{equation}
\log\mu_{\rm loc}
=\frac12\log\det G_g+\frac12\log\det G_f
+\frac12\log\det G_r
-\frac12\log(\nu^{\mathsf T}\nu)
-\log|\det\Delta_{\rm FP}|.
\label{eq:C-local-measure}
\end{equation}

In the combined local density \(L_{\rm nonPf}=-\tfrac12\log|\det H_{154}|+\log|\det\Delta_{\rm FP}|+\log\mu_{\rm loc}\), the explicit and inverse Faddeev--Popov determinants therefore cancel.

Since every column except \(\nu\) is tangent to the constraint surface, the Schur decomposition of the Gram matrix gives

\begin{equation}
\begin{gathered}
F_\nu:=[\mathsf V_g,L,\nu,h,\mathsf V_{\rm ret}],\qquad
|\det F_\nu|^2
=\det G_g\det G_f(\nu^{\mathsf T}\nu)\det D\det G_r,\\
F_A:=[\mathsf V_g,L,A,h,\mathsf V_{\rm ret}],\qquad
\det F_A=\frac{2q_{\rm fix}}{\nu^{\mathsf T}\nu}\det F_\nu.
\end{gathered}
\label{eq:C-frame-determinant}
\end{equation}

The number of columns is \(24+45+1+154+16=240\), and the radial-column coefficient uses \(\nu^{\mathsf T}A=2q_{\rm fix}\). Substituting these relations and \(\det H_{154}=\det H_c/\det D\) into the local density gives Eq.~\eqref{eq:nonpf-density} in the main text.

Separating the individual gradients with \(J_{B,c}:=-\tfrac12d\log|\det H_c|\), we have, at fixed \(q_{\rm fix}\),

\begin{equation}
J_B=J_{B,c}+\frac12d\log\det D,\qquad
J_{\rm geom}=d\log|\det F_A|-\frac12d\log\det D.
\label{eq:C-individual-gradients}
\end{equation}

The terms from polar decomposition cancel in the sum, and the first derivatives of the remaining determinants vanish under the reflection conditions of the next subsection.

\subsection{Reflection covariance of residual and retained spaces and first-order cancellation}\label{app:C-reflection}

Let \(a\) denote the coefficients of the reference configuration, and choose a timelike spacetime covector annihilating the rank-9 background, with \(n_t^{\mathsf T}\eta n_t=-1\) and \(n_t^{\mathsf T}a=0\). The Lorentz time reflection

\begin{equation}
T_{10}=I+2(\eta n_t)n_t^{\mathsf T},\qquad T=T_{10}\otimes I_{24}
\label{eq:C-time-reflection}
\end{equation}

satisfies \(T^2=I\), \(T^{\mathsf T}\mathsf JT=\mathsf J\), \(|\det T|=1\), \(Ta=a\), and \(T\mathsf O=-\mathsf O\). This discrete time reflection belongs to the Lorentz orthogonal group but not to its identity component; it is used here only for the bosonic local construction. In the rest frame, the temporal components of the boost orbit span the nine-dimensional \(W_A\), and the temporal-deformation columns span its trace-orthogonal complement.

Impose horizontality and Hessian orthogonality on the residual columns through

\begin{equation}
U^{\mathsf T}\mathsf J\mathsf E_h=0,\qquad
\mathsf O^{\mathsf T}H_{\mathcal L}\mathsf E_h=0.
\label{eq:C-horizontal-conditions}
\end{equation}

The first condition removes \(W_A\) from the temporal component of a residual vector, leaving a temporal component of the form \(\mathsf O\beta\). Reflection covariance of the reference Hessian eliminates the even--odd cross block, so the second condition becomes \((\mathsf O^{\mathsf T}H_{\mathcal L}\mathsf O)\beta=0\). The numerically established nondegeneracy of the 15-dimensional temporal Hessian at the reference background then gives \(\beta=0\); thus the chosen residual representatives have no temporal component and satisfy \(T\mathsf E_h=\mathsf E_h\).

Similarly, horizontality of the modulus tangent, \(U^{\mathsf T}\mathsf Jv_u=0\), implies that its reflection-odd component belongs to the temporal-deformation space. Writing \(v_u=v_{u,+}+\mathsf O\gamma\), we obtain

\begin{equation}
T[v_u,\mathsf O]=[v_u,\mathsf O]K_r,
\qquad K_r=\begin{pmatrix}1&0\\-2\gamma&-I_{15}\end{pmatrix},
\qquad |\det K_r|=1.
\label{eq:C-retained-reflection}
\end{equation}

This expresses reflection covariance of the retained subspace.

In the reflection-adapted chart of Eq.~\eqref{eq:A-local-chart}, the residual columns are reflection even and the prescribed temporal columns are reflection odd. Geometric time reflection therefore maps \(A(c,y)\) to \(A(-c,y)\). Invariance of the fixed-scale bosonic action gives \(\mathcal L_{\rm loc}(-c,y)=\mathcal L_{\rm loc}(c,y)\), and hence \(f(-c,y)=f(c,y)\) for the residual stationarity equations defined in Appendix~\ref{app:A-splitting}. Since \(f(c,y(c))=0\), local uniqueness of the residual solution on the section \(u=0\) yields \(y(-c)=y(c)\). Consequently, the residual Hessian in the same coordinate basis and the absolute determinant of the full coordinate frame are even functions of \(c\):

\begin{equation}
H_c(-c)=H_c(c),\qquad |\det F_A(-c)|=|\det F_A(c)|.
\label{eq:C-even-determinants}
\end{equation}

For the second equality, the color-gauge frame and the full Lorentz frame transform by \(T\) and fixed changes of generator basis of unit absolute determinant, while the radial, residual, and retained columns transform as above. Substitution into Eq.~\eqref{eq:nonpf-density} gives the first-order cancellation in all 15 directions expressed in Eq.~\eqref{eq:nonpf-cancellation}.

The cancellation requires a rank-9 background with a timelike kernel, a quotient including the full Lorentz orbit, a horizontal and Hessian-orthogonal residual complement, reflection covariance of the retained space, and nonsingular Hessian, Gram, and full-frame matrices. It concerns the first-order local non-Pfaffian gradient in the original \((c,u)\) chart. The required nonsingularity of the nonzero-curvature Hessian, residual block, and local coordinate matrices is supported numerically at the reference background. Appendix~\ref{app:B} provides independent rigorous results within the static reflection sector. The reflection properties of a single-chirality Pfaffian and the global integration cycle remain separate questions.

\subsection{Agreement between the 169-direction Gaussian and the 154+15 splitting}\label{app:C-gaussian}

We compare the two descriptions using the same background, physical quotient, modulus coordinate, Hessian including the constraint force, and Fresnel prescription. Let \(\mathsf T\) denote the invertible map from the 154 residual and 15 temporal-deformation coordinates to an orthonormal basis of the 169-dimensional nonzero-curvature space. In these split coordinates, the Hessian is

\begin{equation}
\begin{gathered}
\widetilde H=\mathsf T^{\mathsf T}H_{169}\mathsf T
=\begin{pmatrix}H_{hh}&H_{hc}\\H_{ch}&H_{cc}\end{pmatrix},
\qquad h_o=H_{cc}-H_{ch}H_{hh}^{-1}H_{hc},\\
\det\widetilde H=\det H_{hh}\det h_o,\qquad
\operatorname{sig}\widetilde H=\operatorname{sig}H_{hh}+\operatorname{sig}h_o.
\end{gathered}
\label{eq:C-schur-gaussian}
\end{equation}

At the reference point, the Hessian-adapted splitting gives \(H_{hc}=H_{ch}=0\), so \(h_o=H_{cc}\). We retain the Schur-complement form to display the determinant and signature relations.

We assume \(H_{hh}\) and \(h_o\) are nondegenerate and retain the volume-element factor \(|\det\mathsf T|\) together with \(\det\widetilde H=(\det\mathsf T)^2\det H_{169}\).

With the fixed-scale differential \(j_b\) and quotient projector \(P\) of Appendix~\ref{app:C-measure}, let \(t_u:=Pj_b(v_u)\) be the physical modulus tangent and set \(s_u:=\|t_u\|_+>0\). Define the 169-column frame \(X_{169}:=[Ph,\,Pj_b(\mathsf O)]\), comprising 154 residual and 15 temporal-deformation columns on the physical quotient. Let \(U_{169}\) be an orthonormal basis for the complement of \(t_u\) in the 170-dimensional physical quotient, so \(H_{169}=U_{169}^{\mathsf T}\mathscr HU_{169}\). The split-to-orthonormal coordinate map is \(\mathsf T=U_{169}^{\mathsf T}X_{169}\). Since \(\mathscr Ht_u=0\), it follows that \(\widetilde H=X_{169}^{\mathsf T}\mathscr HX_{169}\) and \((\det\mathsf T)^2=\det\Gamma\), where \(\Gamma\) is defined below. Thus,

\begin{equation}
\begin{gathered}
\Gamma=X_{169}^{\mathsf T}\left(I-\frac{t_ut_u^{\mathsf T}}{s_u^2}\right)X_{169},\\
\det\operatorname{Gram}[X_{169},t_u]=s_u^2\det\Gamma,\qquad
\det(X_{169}^{\mathsf T}\mathscr HX_{169})=\det\Gamma\det H_{169}.
\end{gathered}
\label{eq:C-modulus-gram}
\end{equation}

Here \(H_{154}\) is obtained by normalizing the residual columns through polar decomposition, while \(G_r\) is the retained Gram matrix for one modulus and 15 temporal directions after Schur elimination of the residuals. In particular, \(\det\operatorname{Gram}[X_{169},t_u]=\det D\,\det G_r\), while \(H_{hh}=H_c\) and \(\det H_{154}=\det H_c/\det D\). Together with Eqs.~\eqref{eq:C-schur-gaussian} and~\eqref{eq:C-modulus-gram}, these identities give

\begin{equation}
\frac{\sqrt{\det G_r}}
{\sqrt{|\det H_{154}|}\sqrt{|\det h_o|}}
=\frac{s_u}{\sqrt{|\det H_{169}|}}.
\label{eq:C-gaussian-density}
\end{equation}

The residual Gram factor is already included in the normalization of \(H_{154}\); multiplying by it again would double-count it. The form \(h_o\) is expressed in the same 15 coefficients, and changing these coefficients requires the corresponding Jacobian to change as well.

By Eq.~\eqref{eq:C-fresnel} and additivity of the signature, the product of the 154- and 15-direction Fresnel phases also agrees with the 169-direction phase. With the reference-point Pfaffian, Faddeev--Popov, and coarea factors treated in the same convention, integrating the 15 retained directions in the reference-point quadratic Fresnel approximation reproduces the leading local 169-direction Gaussian density and phase. In the deformation-dependent analysis of this paper, the 15 coefficients remain unintegrated parameters.

\section{Identities for the fermionic color--spin coupling vector and eigenvalue sensitivity}\label{app:D}

\subsection{Double commutator and sign of the color--spin coupling vector}\label{app:D-source}

Let \(e_\ell\) be trace-orthonormal traceless Hermitian color matrices. Define \(D_xY=i[x,Y]\), \(D_i=D_{A_i^{(0)}}\), and \(C_\ell=D_{e_\ell}\). On the real color-coefficient space, \(D_x\) is antisymmetric and \([D_x,D_y]=D_{D_xy}\). For the rest-frame operators \(\mathcal K_F=i\sum_i\widehat\gamma^i\otimes D_i\) and \(\mathcal J_\ell=iI_{16}\otimes C_\ell\), define the spin-dependent commutator operator by

\begin{equation}
\mathcal R_\ell:=-i\sum_{1\le i<j\le9}
\widehat\gamma^i\widehat\gamma^j\otimes
\left(\{D_i,[D_j,C_\ell]\}-\{D_j,[D_i,C_\ell]\}\right).
\label{eq:D-source-operator}
\end{equation}

For \(i\ne j\), \(\widehat\gamma^i\widehat\gamma^j\) is real antisymmetric and the color anticommutators are real symmetric. Thus \(\mathcal R_\ell\) is Hermitian, with its sign fixed by the double-commutator expansion. Expanding the double commutator using the spatial Clifford relations, the terms with equal spin indices give

\begin{equation}
-iI_{16}\otimes\sum_{i=1}^{9}[D_i,[D_i,C_\ell]]
=iI_{16}\otimes D_{-\sum_iD_i^2e_\ell}
=\mathcal J_{\Delta e_\ell}.
\label{eq:D-equal-spin}
\end{equation}

Combining the two ordered terms with distinct indices gives \(\mathcal R_\ell\). This is the operator identity in Eq.~\eqref{eq:color-spin-double-commutator}; the expansion itself does not use classical stationarity.

If \(P_r\) is a spectral projector of the full \(\mathcal K_F\), then \([P_r,\mathcal K_F]=0\). Cyclicity of the trace gives

\begin{equation}
\operatorname{Tr}\!\left(P_r[\mathcal K_F,[\mathcal K_F,\mathcal J_\ell]]\right)
=\operatorname{Tr}\!\left([\mathcal K_F,[\mathcal K_F,P_r]]\mathcal J_\ell\right)=0.
\label{eq:D-cyclic-trace}
\end{equation}

As \(\Delta\) is real symmetric, \(d_r^{-1}\operatorname{Tr}(P_r\mathcal J_{\Delta e_\ell})=(\Delta q_r)_\ell\). Together with \(s_{r,\ell}=-d_r^{-1}\operatorname{Tr}(P_r\mathcal R_\ell)\), Eq.~\eqref{eq:D-cyclic-trace} gives \((\Delta q_r)_\ell-s_{r,\ell}=0\). Here \(P_r\) is a spectral projector of the full \(\mathcal K_F\).

\subsection{Projection onto physical temporal deformations and reconstruction of sensitivity}\label{app:D-sensitivity}

Let \(W_A=\operatorname{span}_{\mathbb R}\{A_1^{(0)},\ldots,A_9^{(0)}\}\), which is nine-dimensional at the selected reference background. The fixed-\(Q_\rho\) stationarity equations give \(W_A\subseteq\ker(\Delta-2\lambda_QI)\). Since \(\Delta\) is real symmetric, \(W_A^\perp\) is invariant.

Using the isometric embedding \(\mathcal E\) in the same trace coefficient metric, we have

\begin{equation}
\mathcal E^{\mathsf T}\mathcal E=I_{15},\qquad
\mathcal E^{\mathsf T}W_A=0,\qquad
\mathcal E^{\mathsf T}\Delta W_A=0
\label{eq:D-color-projection}
\end{equation}
and \(\mathcal E^{\mathsf T}\Delta\mathcal E\) is given by Eq.~\eqref{eq:A-time-hessian}. Since the conditional background satisfies \(D_cy(0)=0\) and \(D_c\beta(0)=0\), the first-order fermionic variation is \(\partial_{c^a}\mathcal K_F|_0=\mathcal J_{H_a}=\sum_\ell\mathcal E_{\ell a}\mathcal J_\ell\). The sensitivity in Eq.~\eqref{eq:eigenvalue-sensitivity} therefore equals \(a_r=\mathcal E^{\mathsf T}q_r\).

As \(\operatorname{im}\mathcal E=W_A^\perp\), projection of \(\Delta q_r=s_r\) gives \((2\lambda_QI_{15}-H_{\rm time})a_r=s_{r,\mathcal E}\) without inversion. For any traceless Hermitian color matrix \(X\), \(\operatorname{Tr}(X\Delta X)=\sum_{i=1}^{9}\|[A_i^{(0)},X]\|_F^2\). At the reference background with \(A_0^{(0)}=0\), the full-rank (rank 24) color-gauge tangent map implies that no such nonzero \(X\) commutes with every spatial matrix. Under this numerically checked rank condition, \(\Delta\), and hence its restriction \(\mathcal E^{\mathsf T}\Delta\mathcal E=2\lambda_QI_{15}-H_{\rm time}\), is positive definite. Thus the inverse used in Section~\ref{sec:source} exists at the selected background; nondegeneracy of \(H_{\rm time}\) for the conditional classical response is a separate condition. The color--spin coupling vector uses the full spectral projector; its two-plane concentration for the lowest doublet is numerical.

Under the exact block-structure hypothesis of Section~\ref{sec:plane} and Eq.~\eqref{eq:A-color-blocks}, choose the reflection-adapted color and spatial bases so that \(R=\operatorname{diag}(-1,I_4)\) and \(O=\operatorname{diag}(-I_3,I_6)\), where the first three spatial directions are reflection odd. With the real symmetric spatial Clifford matrices used here, set \(S=\widehat\gamma^1\widehat\gamma^2\widehat\gamma^3\). Then \(S^{\mathsf T}=-S\), \(S^2=-I_{16}\), and \(S^\dagger S=I_{16}\). Writing \(\mathsf C_R\) for the real orthogonal adjoint action of \(R\) on color coefficients, these choices give the color--spin covariance relations

\begin{equation}
\mathsf C_RD_i\mathsf C_R^{-1}=\sum_jO_{ij}D_j,
\qquad S\widehat\gamma^iS^{-1}=-\sum_jO_{ij}\widehat\gamma^j.
\label{eq:D-color-spin-covariance}
\end{equation}

Let \(\mathscr C\) be complex conjugation in these real coefficient bases, and define \(\Theta=(S\otimes\mathsf C_R)\mathscr C\). Since \(S\) and \(\mathsf C_R\) are real and \(\mathsf C_R^2=I_{24}\), we have \(\Theta^2=-I_{384}\). At the reference configuration with \(A_0^{(0)}=0\), complex conjugation and conjugation by \(S\otimes\mathsf C_R\) each reverse the sign of \(\mathcal K_F=i\mathcal M_F[A^{(0)}]\); hence \(\Theta\mathcal K_F\Theta^{-1}=\mathcal K_F\). The same covariance gives \(\Theta\mathcal J_v\Theta^{-1}=-\mathcal J_{\mathsf C_Rv}\). Since \([\mathsf C_R,\Delta]=0\), transforming the double commutator in Eq.~\eqref{eq:color-spin-double-commutator} gives

\begin{equation}
\Theta\mathcal R_v\Theta^{-1}=-\mathcal R_{\mathsf C_Rv}.
\label{eq:D-reflection-source-law}
\end{equation}

Because \(\Theta\) commutes with \(\mathcal K_F\), every complete eigenspace, and every direct sum of complete eigenspaces, has a \(\Theta\)-invariant orthogonal spectral projector: \(\Theta P_r\Theta^{-1}=P_r\). The transformation law for \(\mathcal R_v\) in Eq.~\eqref{eq:D-reflection-source-law}, together with the reality of the trace for Hermitian operators and the definition of the color--spin coupling vector, implies \(s_r(\mathsf C_Rv)=-s_r(v)\). Thus the projected vector \(s_{r,\mathcal E}\) lies in the five-real-dimensional reflection-odd sector of the 15-dimensional physical temporal-deformation space: the \(1+4\) color decomposition has eight real reflection-odd directions, three of them in \(W_A\), leaving five in \(W_A^\perp\). Moreover, \(\Theta^2=-I_{384}\) implies that every exact eigenvalue of \(\mathcal K_F\) has even multiplicity. These are analytic consequences of the exact block structure and apply to complete spectral projectors on the certified branch through \(A_*\). At the numerical reference background, these symmetry relations hold to numerical precision. This color--spin symmetry is distinct from geometric time reflection and physical CPT. The strong concentration of the coupling vectors in the two-plane \(P_0\) within the odd five-dimensional space is a separate numerical result, not enforced by the reflection symmetry.

\subsection{Eigenvalue clusters and numerical consistency checks}\label{app:D-numerics}

Under exact reflection, the antiunitary symmetry of Appendix~\ref{app:D-sensitivity} proves that every eigenvalue of \(\mathcal K_F\) has even multiplicity. Consistently with this structure, sorting the 192 positive numerical eigenvalues and grouping neighbors with a difference threshold of \(10^{-8}\max(1,|\lambda|)\) gives 89 groups: 88 doublets and one 16-state cluster. This numerical classification does not assert that the 16-state cluster is a single exactly degenerate eigenvalue. Each projector \(P_r=U_rU_r^\dagger\), constructed from orthonormal eigenvectors of the group, is independent of basis rotations within it. The mean eigenvalues of the lowest and next groups are approximately 0.0682001313 and 0.4394064314, respectively.

The positive spectrum has a maximum within-cluster width of approximately \(1.24\times10^{-14}\) and a maximum relative width of approximately \(2.28\times10^{-14}\). The relative width is the maximum-minus-minimum eigenvalue divided by the group's arithmetic mean. For an exactly degenerate group, Eq.~\eqref{eq:pfaffian-sum} gives the multiplicity-weighted mean sensitivity; for a general spectral group, \(\operatorname{Tr}(P_r\mathcal K_F^{-1}\partial_{c^a}\mathcal K_F)\) is its exact contribution in the positive-spectrum convention used here. The sum \(J_{\rm spec}\) over the 89 numerical groups agrees with the direct inverse-matrix-trace gradient \(J_{\rm dir}\), evaluated in the same rest frame and 15 temporal directions, to \(\|J_{\rm spec}-J_{\rm dir}\|_2/\|J_{\rm dir}\|_2\simeq1.49\times10^{-14}\).

The fractions in the main text involve different spaces and normalizations and are distinguished below. The projector \(\Pi_0\) is the orthogonal projection onto the two-plane within the 15-dimensional temporal-deformation space.

\begin{center}
\normalsize
\renewcommand{\arraystretch}{1.15}
\begin{tabular}{>{\raggedright\arraybackslash}p{0.28\linewidth}>{\raggedright\arraybackslash}p{0.46\linewidth}r}
\toprule
Quantity & Definition and denominator & Value (approx.)\\
\midrule
Auxiliary zero-mode subspace weight & $\tfrac12\operatorname{Tr}(P_*P_Z)$: doublet average & 0.2995\\
Color--spin vector plane fraction & $\|\Pi_0s_{*,\mathcal E}\|^2/\|s_{*,\mathcal E}\|^2$ & 0.9835\\
Eigenvalue-sensitivity plane fraction & $\|\Pi_0a_*\|^2/\|a_*\|^2$ & 0.9943\\
Lowest-cluster signed projection & $J_*\!\cdot J_{\rm Pf}/\|J_{\rm Pf}\|^2$ & 0.9862\\
Relative residual of the lowest-cluster approximation & $\|J_{\rm Pf}-J_*\|/\|J_{\rm Pf}\|$ & 0.0138\\
Full one-loop-gradient plane fraction & $\|\Pi_0J_{\rm1L}\|^2/\|J_{\rm1L}\|^2$ & 0.9941\\
\bottomrule
\end{tabular}
\end{center}
\medskip

The auxiliary zero-mode subspace weight is defined in state space, whereas the plane fractions are evaluated in the 15-dimensional temporal-deformation space. The signed projection and relative residual are not complementary quantities. The full one-loop gradient uses the conditional first-order cancellation of Appendix~\ref{app:C-reflection}. Transporting these numerical eigenspaces and fractions to \(A_*\) requires controlled operator errors, isolation gaps, and projector perturbation bounds in addition to classical pointwise proximity.

\section{Conditional classical spatial response}\label{app:E}

\subsection{Angle independence of the quadratic response from a Pauli pair}\label{app:E-pauli}

As in Section~\ref{sec:spatial}, let \(e\) be the unit vector of the one-dimensional block in the \(1+4\) decomposition, and let \(t\) be the normalized complex-kernel vector embedded in the five-dimensional color space, with \(e^\dagger t=0\). The same trace-orthonormal Hermitian matrices as in Eq.~\eqref{eq:soft-plane-basis} form a basis of the plane:

\begin{equation}
H_1=\frac{et^\dagger+te^\dagger}{\sqrt2},
\qquad
H_2=\frac{-iet^\dagger+ite^\dagger}{\sqrt2}.
\label{eq:E-pauli-basis}
\end{equation}

Setting \(\Pi_{\rm c}=ee^\dagger+tt^\dagger\), we have \(H_1^2=H_2^2=\Pi_{\rm c}/2\) and \(\{H_1,H_2\}=0\), so the square of \(H(\varphi)=H_1\cos\varphi+H_2\sin\varphi\) is angle independent.

For each spatial matrix \(A_i^{(0)}\), Hermiticity and the two trace-orthogonality conditions \(\operatorname{Tr}(H_aA_i^{(0)})=0\) (\(a=1,2\)) imply \(e^\dagger A_i^{(0)}t=0\). Consequently,

\begin{equation}
\begin{aligned}
H(\varphi)A_i^{(0)}H(\varphi)
&=\frac12\bigl[(t^\dagger A_i^{(0)}t)ee^\dagger
+(e^\dagger A_i^{(0)}e)tt^\dagger\bigr],\\
[H(\varphi),[H(\varphi),A_i^{(0)}]]
&=\frac12\{\Pi_{\rm c},A_i^{(0)}\}
-2H(\varphi)A_i^{(0)}H(\varphi).
\end{aligned}
\label{eq:E-pauli-source}
\end{equation}

Both expressions are independent of \(\varphi\). The quadratic normalization to fixed \(Q_\rho\) is common to all directions because \(\operatorname{Tr}H(\varphi)^2=1\). At quadratic order, the 154 residual stationarity equations are driven by the double-commutator terms and the fixed-scale normalization contribution, both independent of \(\varphi\). Since the same invertible residual Hessian governs all unit directions in the plane, the induced quadratic spatial response, and hence \(R_2\) and \(g_2\), are independent of \(\varphi\). Positivity of the coefficients is the numerical result for the selected background in Eq.~\eqref{eq:spatial-coefficients}.

\subsection{Residual response and variation of the spatial projector}\label{app:E-spatial}

On the fixed-scale section, \(\beta^2=1+\rho^2+O(\rho^4)\) multiplies all \(K\)-eigenvalues equally. It cancels from the hierarchy ratio but contributes the reference gap \(\Delta_0\) to \(g_2\). Since \(G_{0i}=O(r^3)\), temporal--spatial mixing has no quadratic contribution; the remaining response comes from classical stationarity in the 154 residual directions.

Fix the Lorentz rest frame of the reference background and its spatial eigenbasis, and let \(P_3^{\rm sp}\) be the orthogonal projector onto the three largest eigenvalues of the nine-dimensional real symmetric Gram matrix \(G_{\rm sp}\). The finite gap between the third and fourth eigenvalues at the reference point allows this leading three-dimensional subspace to be tracked independently of within-group degeneracies. Its expansion is

\begin{equation}
P_3^{\rm sp}(\rho,d)=P_{3,0}^{\rm sp}
+\rho^2P_{3,2}^{\rm sp}(d)+O(\rho^4).
\label{eq:E-spatial-projector}
\end{equation}

Writing \(G_{\rm sp}(\rho,d)=G_{{\rm sp},0}+\rho^2 C_{\rm sp}(d)+O(\rho^4)\) and ordering the reference spatial eigenvalues \(\lambda_i^{\rm sp}\) from largest to smallest, the quadratic projector coefficient is

\begin{equation}
(P_{3,2}^{\rm sp})_{ij}=
\frac{(C_{\rm sp})_{ij}}{\lambda_i^{\rm sp}-\lambda_j^{\rm sp}}
\quad(i\leq3<j),\qquad
(P_{3,2}^{\rm sp})_{ji}=(P_{3,2}^{\rm sp})_{ij}.
\end{equation}

The within-group blocks vanish by \(P^2=P\), and \([G_{\rm sp},P]=0\) determines the cross-group blocks. Only cross-group gaps appear in the denominators, so within-group degeneracies do not obstruct the expansion. Along the projected-gradient direction in the distinguished plane, \(\|P_{3,2}^{\rm sp}\|_F\simeq1.0773\), showing mixing between the two groups in the fixed reference spatial frame.

The quadratic coefficients are obtained by substituting \(y(rd)=\tfrac12r^2y_{,ab}(0)d^ad^b+O(r^4)\) into the spatial Gram matrix and normalizing with \(r^2=q_0\rho^2\). The coefficient \(R_2\) is the quadratic coefficient of the ratio between the means of the leading three and lower six eigenvalues. At an exact reflection-symmetric stationary solution, the odd-sector equation implies \([G_Z^{\rm R},C_Z]=0\) (Eq.~\eqref{eq:B-commuting-gram}). The certified bound \(b^{\mathsf T}b>0\) ensures that \(C_Z\neq0\). A real symmetric matrix commuting with this antisymmetric matrix has the form \(G_Z^{\rm R}=g_\perp(I_3-P_{\rm ax})+g_\parallel P_{\rm ax}\), with \(P_{\rm ax}\) defined in Eq.~\eqref{eq:B-axis-projector}. Thus the eigenvalue on the plane perpendicular to \(b\) is exactly repeated. The reflection-odd spatial Gram block is \(2G_Z^{\rm R}/5\), and reflection symmetry makes its mixed entries with the even block vanish. This eigenvalue therefore has multiplicity at least two in the full spatial Gram matrix at the stationary solution.

For the certified nearby exact stationary solution, this doublet is precisely the fourth--fifth pair in the descending spatial spectrum. Rational spectral enclosures for the reference center, combined with the rigorous proximity bound in the same reflection-adapted rest frame, give \(\lambda_3^{\rm sp}-\lambda_4^{\rm sp}>0.09902281658\) and \(\lambda_5^{\rm sp}-\lambda_6^{\rm sp}>0.01183653172\), so the doublet is isolated from the other seven eigenvalues. At the numerical reference background, the corresponding pair likewise occupies the fourth and fifth positions in the ordered positive spatial spectrum, within numerical resolution. The numerical evaluation of \(g_2\) treats this unresolved pair as degenerate and uses the larger eigenvalue of \(C_{\rm sp}\) restricted to its two-dimensional subspace. This gives the ordered fourth-eigenvalue quadratic shift at an exactly degenerate base point; at a fixed numerical background with a nonzero splitting, the strict \(\rho\to0\) coefficient can differ. The coefficient \(g_2\) remains a numerical result at the reference background; an error-controlled value at the nearby exact solution would require separate bounds on the conditional response.

\section*{Computational implementation and figure preparation}\phantomsection\labelunnumbered{sec:comp-impl}
\addcontentsline{toc}{section}{Computational implementation and figure preparation}

All six panels in Figures~\ref{fig:temporal-eigenvalue}--\ref{fig:spatial-response} were prepared from previously computed numerical data using Python and Matplotlib. OpenAI's ChatGPT (GPT-6) assisted in developing and refining plotting scripts for selected panels. The plotted values were compared with the underlying numerical results.

OpenAI's ChatGPT also assisted in developing and revising selected computational code used for the numerical analyses in this work.

Appendix~\ref{app:B} presents a computer-assisted existence proof for a nearby exact constrained classical solution, together with rigorous bounds in the static reflection sector. The fermionic spectra and Pfaffian responses are evaluated separately at the numerical reference background.

\section*{Reproducibility and data availability}\phantomsection\labelunnumbered{sec:data}
\addcontentsline{toc}{section}{Reproducibility and data availability}

Data and plotting scripts for Figures~\ref{fig:temporal-eigenvalue}--\ref{fig:spatial-response}, fixed-\(Q_\rho\) classical existence and rigorous pointwise proximity certificates for 23 configurations, saved Pfaffian spectra and verification scripts, rigorous bounds on the fourth--fifth spatial eigenvalue ordering, and conditional spatial-response tensors are available at \url{https://doi.org/10.5281/zenodo.23997377}. The classical certificates concern a nearby exact stationary solution; the numerical responses use the reference background. The scripts check saved arrays, not the full original calculations. See Ref.~\cite{muramatsu2026} for the background.

\section*{Declaration of generative AI and AI-assisted technologies in the manuscript preparation process}\phantomsection\labelunnumbered{sec:ai-declaration}
\addcontentsline{toc}{section}{Declaration of generative AI and AI-assisted technologies in the manuscript preparation process}

During the preparation of this work, the author used OpenAI's ChatGPT to assist with computational code generation and revision, manuscript organization, bibliographic verification, and English-language editing.

The author reviewed and edited AI-assisted material, made the final scientific and editorial decisions, and takes full responsibility for the content of the manuscript and the reported results.

\begingroup
\normalsize\setstretch{1.12}
\setlength{\parindent}{0pt}

\endgroup
\end{document}